\documentclass[10pt]{article}
\usepackage[pdftex]{graphicx}
\usepackage{caption}
\usepackage{subcaption}
\usepackage{dcolumn}
\usepackage{bm}
\usepackage{color}
\usepackage{amsfonts}
\usepackage{amsmath}
\usepackage{stmaryrd}
\usepackage{epsfig}
\usepackage[english]{babel}
\usepackage[all]{xy}
\usepackage{ulem}
\usepackage[top=25.4mm, bottom=25.4mm, left=31.7mm, right=32.2mm]{geometry}

\input{tcilatex}
\begin{document}

\title{\textbf{Second-order integrable Lagrangians and WDVV equations}}
\author{E.V. Ferapontov$^{1}$, M.V. Pavlov$^{2}$, Lingling Xue$^{3}$}
\date{}
\maketitle

\renewcommand{\baselinestretch}{1.25}

\vspace{-5mm}

\begin{center}
$^{1}$Department of Mathematical Sciences \\[0pt]
Loughborough University \\[0pt]
Loughborough, Leicestershire LE11 3TU, UK \\[0pt]
\ \\[0pt]
$^{2}$Lebedev Physical Institute \\[0pt]
Russian Academy of Sciences,\\[0pt]
Leninskij Prospekt 53, 119991 Moscow, Russia\\[0pt]
\ \\[0pt]
$^{3}$Department of Mathematics\\[0pt]
Ningbo University\\[0pt]
Ningbo 315211, P.R. China \\[0pt]
\ \\[0pt]
e-mails: \\[1ex]
\texttt{E.V.Ferapontov@lboro.ac.uk}\\[0pt]
\texttt{mpavlov@itp.ac.ru}

\texttt{xuelingling@nbu.edu.cn}\\[0pt]
\end{center}

\bigskip

\begin{abstract}
We investigate integrability of Euler-Lagrange equations associated with 2D
 second-order Lagrangians of the form
\begin{equation*}
\int f(u_{xx},u_{xy},u_{yy})\ dxdy.
\end{equation*}
By deriving integrability conditions for the Lagrangian density $f$,
examples of integrable Lagrangians expressible via elementary functions,
Jacobi theta functions and dilogarithms  are constructed. A link of  second-order integrable Lagrangians to  WDVV equations is established. Generalisations to 3D second-order integrable Lagrangians are also discussed.
\end{abstract}

\noindent MSC: 35Q51, 37K05, 37K10, 37K20, 53D45.

\bigskip

\noindent \textbf{Keywords:} {Second-order Lagrangians, systems of
hydrodynamic type, integrability (diagonalisability) conditions, Jacobi theta
functions, Chazy equation, WDVV equations.}

\newpage

\bigskip

\centerline{\it Dedicated to the memory of  Boris Anatolievich Dubrovin}

\tableofcontents


\section{Introduction and summary of the main results}

We investigate second-order Lagrangians 
\begin{equation}
\int f(u_{xx},u_{xy},u_{yy})\ dxdy,  \label{2}
\end{equation}
such that the corresponding Euler-Lagrange equations are integrable (in the sense to be explained below). 
Examples of integrable Lagrangians (\ref{2}) have appeared in the
mathematical physics literature, thus, the  Lagrangian density
\begin{equation}\label{Mir}
f=u_{xy}(u_{xx}^2-u_{yy}^2)+\alpha (u_{xx}^2-u_{yy}^2)+u_{xy}(\beta
u_{xx}+\gamma u_{yy})
\end{equation}
governs integrable geodesic flows on a 2-torus which possess a fourth-order
integral polynomial in the momenta \cite{BM}. Similarly,  the density
\begin{equation}\label{dKP}
f=u_{yy}^{2}+u_{xx}^{2}u_{yy}+u_{xx}u_{xy}^{2}+\frac{1}{4}u_{xx}^{4}
\end{equation}
governs integrable Newtonian equations possessing a fifth-order polynomial integral.
In Section \ref{sec:int2D} we investigate the integrability aspects of 2D  Lagrangians (\ref{2}). Our main results  can be summarised as follows.

\begin{itemize}

\item The Euler-Lagrange equation coming from Lagrangian (\ref{2}) can be represented as a four-component Hamiltonian system of hydrodynamic type  (Section \ref{sec:h2}). The requirement of its hydrodynamic integrability (which is equivalent to the vanishing of the corresponding Haantjes tensor) leads to an involutive  system of third-order PDEs for the Lagrangian density $f$ (Section \ref{sec:i2}). Analysis of the integrability conditions reveals that integrable Lagrangians (\ref{2}) locally depend on six arbitrary functions of one variable. Furthermore, the integrability conditions are themselves integrable -- a standard phenomenon in the theory of integrable systems. 

\item The class of integrable Lagrangians  (\ref{2}) is invariant under the symplectic group ${\rm Sp}(4, \mathbb{R})$; under this action the Lagrangian density $f$ transforms as a  genus two Siegel modular form of weight $-1$ (Section \ref{sec:equiv}). In particular, the integrability conditions  can be represented via ${\rm Sp}$-invariant operations known  as generalised Rankin-Cohen (Eholzer-Ibukiyama) brackets (Section \ref{sec:RC}).

\item  Potentials $U(x, t)$ of classical Newtonian equations $\ddot{x}=-U_{x}$ that possess a fifth-order polynomial integral are governed by a Lagrangian (\ref{2})  with density (\ref{dKP}) (Section \ref{sec:1/2}).

\item Integrable Lagrangians (\ref{2}) are related to WDVV prepotentials of the form
$$
F(t_{1},t_{2},t_{3},t_{4})=\frac{1}{2}t_{1}^{2}t_{4}+t_{1}t_{2}t_{3}+W(t_{2},t_{3},t_{4});
$$
here $W$ is a partial Legendre transform of the Lagrangian density $f$ (Section \ref{sec:i0}). This correspondence works both ways: using known solutions of WDVV equations one can construct new integrable Lagrangians (\ref{2}). Conversely, integrable Lagrangian densities $f$ give rise to WDVV prepotentials. Examples  of this kind are given in Sections \ref{sec:WDVVex} and \ref{sec:LWDVV}. 

\item Further examples of integrable Lagrangians (\ref{2}) expressible via elementary functions,  Jacobi theta functions and dilogaritms are constructed in Section \ref{sec:ex}.

\end{itemize}

In Section \ref{sec:int3D} we investigate 3D second-order Lagrangians of the form
\begin{equation}  \label{3}
\int f(u_{xx}, u_{xy}, u_{xt}, u_{yy}, u_{yt}, u_{tt})\ dxdydt.
\end{equation}
Our results can be summarised as follows:
\begin{itemize}
\item Integrable Lagrangians in 3D are governed by a  third-order PDE system for the Lagrangian density $f$ which comes from the requirement that all travelling wave reductions of 3D Lagrangians to 2D are integrable in the sense of Section \ref{sec:int2D} (Section \ref{sec:int3}). 

\item The class of integrable Lagrangians  (\ref{3}) is invariant under the symplectic group ${\rm Sp}(6, \mathbb{R})$; the Lagrangian density $f$ transforms as a  genus three Siegel modular form  of weight $-1$ (Section \ref{sec:equiv3}).

\item Examples of integrable Lagrangians (\ref{3}) are constructed in Section \ref{sec:ex3}. These include the densities
$$
f=u_{yy}^{2}-u_{xx}u_{xt}+u_{xx}^{2}u_{yy}+u_{xx}u_{xy}^{2}+\frac{1}{4}u_{xx}^{4},
$$
$$
f=  (u_{xy}-u_{tt}-u_{xx}u_{xt}+\frac{1}{3}u_{xx}^{3}) ^{3/2},
$$
$$
f=u_{xt}^{-2}(u_{xt}u _{yt}-u_{xx}u_{xt}^2)^{3/2},
$$
coming from the theory of dispersionless KP hierarchy (Section \ref{sec:dKP}).  

\item Classification of integrable densities of the form
$$
f=f(u_{xy}, u_{xt}, u_{yt})
$$
is given in Section \ref{sec:Lob}. Here the generic case is quite non-trivial, involving spherical trigonometry and Schl\"afly-type formulae, and is expressed in terms of the Lobachevky function $L(s)=-\int_0^s\ln  \cos \xi \ d\xi$. 

\end{itemize}

In Section \ref{sec:deform} we discuss examples of integrable dispersive deformations of integrable Lagrangian densities  
(\ref{3}). The general problem of constructing such deformations is largely open.

\medskip

Finally, we recall that paper \cite{FKT}  gives a characterisation of  3D first-order integrable Lagrangians of the form
$$
\int f(u_x, u_y, u_t)\ dx dy dt.
$$
It was pointed out in \cite{FO} that the generic integrable Lagrangian density of this type is an automorphic function of its arguments. Note that  2D first-order Lagrangians  densities $f(u_x, u_y)$ lead to linearisable Euler-Lagrange equations and, therefore, are automatically integrable. On the contrary, for second-order Lagrangian densities $f(u_{xx},u_{xy},u_{yy})$,  the 2D case (\ref{2}) is already nontrivial.

\section{Integrable Lagrangians in 2D}
\label{sec:int2D}

In this section we consider second-order integrable Lagrangians of type (\ref{2}), 
$$
\int f(u_{xx},u_{xy},u_{yy})\ dxdy.  
$$

\subsection{Hydrodynamic form of Euler-Lagrange equations}

\label{sec:h2}

The Euler-Lagrange equation corresponding to Lagrangian (\ref{2}) is a fourth-order PDE
for $u(x,y)$:
\begin{equation}
\left( \frac{\partial f}{\partial u_{xx}}\right) _{xx}+\left( \frac{\partial
f}{\partial u_{xy}}\right) _{xy}+\left( \frac{\partial f}{\partial u_{yy}}
\right) _{yy}=0.  \label{Euler}
\end{equation}
Setting $a=u_{xx}$, $b=u_{xy}$, $c=u_{yy}$ we can rewrite (\ref{Euler}) in
the form
\begin{equation}
b_{x}=a_{y},\quad c_{x}=b_{y},\quad \left( f_{a}\right) _{xx}+\left(
f_{b}\right) _{xy}+\left( f_{c}\right) _{yy}=0.  \label{abc}
\end{equation}
Introducing the auxiliary variable $p$ via the relations
\begin{equation*}
p_{y}=-\left( f_{a}\right) _{x},\text{ \ }p_{x}=\left( f_{b}\right)
_{x}+\left( f_{c}\right) _{y},
\end{equation*}
we can rewrite (\ref{abc}) as a first-order four-component conservative system
\begin{equation}
a_{y}=b_{x},\text{ \ }b_{y}=c_{x},\text{ \ }(f_{c})_{y}=(p-f_{b})_{x},\text{
\ }p_{y}=-(f_{a})_{x}  \label{4}
\end{equation}
or, in matrix form,
\begin{equation*}
{\bf R}w_{y}={\bf S}w_{x}
\end{equation*}
where $w=(a,b,c,p)^{T}$ and
\begin{equation*}
{\bf R}=\left(
\begin{array}{cccc}
1 & 0 & 0 & 0 \\
0 & 1 & 0 & 0 \\
f_{ac} & f_{bc} & f_{cc} & 0 \\
0 & 0 & 0 & 1
\end{array}
\right) ,\quad {\bf S}=\left(
\begin{array}{cccc}
0 & 1 & 0 & 0 \\
0 & 0 & 1 & 0 \\
-f_{ab} & -f_{bb} & -f_{bc} & 1 \\
-f_{aa} & -f_{ab} & -f_{ac} & 0
\end{array}
\right) .
\end{equation*}
Assuming $f_{cc}\neq 0$ we obtain a four-component system of hydrodynamic
type,
\begin{equation}
w_{y}={\bf V}(w)w_{x},\quad {\bf V}(w)={\bf R}^{-1}{\bf S}.  \label{V}
\end{equation}

\medskip

\noindent \textbf{Remark 1.} System (\ref{4}) can be put into a Hamiltonian form. For that
purpose we introduce the new dependent variables $(A,B,C)$ which are related
to $(a,b,c)$ via partial Legendre transform,
\begin{equation*}
A=a,\ B=b,\ C=f_{c},\ h=cf_{c}-f,\ h_{A}=-f_{a},\ h_{B}=-f_{b},\ h_{C}=c.
\end{equation*}
In the new variables, system (\ref{4}) takes the form ($P=p$)
\begin{equation}
A_{y}=B_{x},\text{ \ }B_{y}=(h_{C})_{x},\text{ \ }C_{y}=(P+h_{B})_{x},\text{
\ }P_{y}=(h_{A})_{x},  \label{H}
\end{equation}
which is manifestly Hamiltonian:
\begin{equation*}
\begin{pmatrix}
A \\
B \\
C \\
P
\end{pmatrix}
_{y}=
\begin{pmatrix}
0 & 0 & 0 & 1 \\
0 & 0 & 1 & 0 \\
0 & 1 & 0 & 0 \\
1 & 0 & 0 & 0
\end{pmatrix}
\frac{d}{dx}
\begin{pmatrix}
H_{A} \\
H_{B} \\
H_{C} \\
H_{P}
\end{pmatrix}
,
\end{equation*}
with the Hamiltonian density $H=h(A,B,C)+BP$.

\subsection{Integrability conditions}

\label{sec:i2}

Since hydrodynamic type system (\ref{4}) is conservative, its integrability
by the generalised hodograph method \cite{Tsarev} is equivalent to the
diagonalisability of the corresponding matrix ${\bf V}(w)$ from (\ref{V}). This is
equivalent to the vanishing of the corresponding Haantjes tensor \cite
{Haantjes}. Recall that the Nijenhuis tensor of the matrix $
{\bf V}(w)=(v^{i}_{j}(w))$ is defined as
\begin{equation*}
N_{jk}^i=v^s_{j}\partial_{w^s} v^i_{k}-v^s_{k}\partial_{w^s}
v^i_{j}-v^i_{s}(\partial_{w^j} v^s_{k}-\partial_{w^k} v^s_{j}),
\end{equation*}
where we adopt the notation $w=(a, b, c, p)^T=(w^1, w^2, w^3, w^4)^T$. The
Haantjes tensor is defined as
\begin{equation*}
H^i_{jk}=N^i_{pr}v^p_{j} v^r_{k}+N^p_{jk} v^i_{r}v^r_{p}-N^p_{jr}
v^i_{p}v^r_{k}-N^p_{rk} v^i_{p}v^r_{j}.
\end{equation*}
It is easy to see that both tensors are skew-symmetric in the low indices.
The requirement of vanishing of the Haantjes tensor leads to a system of
 PDEs (integrability conditions) for the Lagrangian density $f(a,b,c)$ which can be represented in symmetric conservative form:
\begin{equation}  \begin{array}{c}\label{int}
(f_{ab}f_{cc}-f_{ac}f_{bc})_a=(f_{bc}f_{aa}-f_{ab}f_{ac})_c,\\
\ \\
(f_{aa}f_{cc}-f_{ac}^2)_a=(f_{aa}f_{bb}-f_{ab}^2)_c,\\
\ \\
(f_{aa}f_{cc}-f_{ac}^2)_c=(f_{cc}f_{bb}-f_{bc}^2)_a,\\
\ \\
(f_{bb}f_{cc}-f_{bc}^2)_b=2(f_{ab}f_{cc}-f_{ac}f_{bc})_c,\\
\ \\
(f_{bb}f_{aa}-f_{ab}^2)_b=2(f_{bc}f_{aa}-f_{ac}f_{ab})_a.\\
\end{array}
\end{equation}
Integrability conditions (\ref{int}) are invariant under the discrete symmetries $a\leftrightarrow c$ and $b\to -b$. Indeed, under the interchange of $a$ and $c$ equation $(\ref{int})_1$ stays the same, while $(\ref{int})_2$, $(\ref{int})_3$ and $(\ref{int})_4$, $(\ref{int})_5$ get interchanged. Strictly speaking, the vanishing of the Haantjes tensor gives only the first four of relations $(\ref{int})$, however, one can show that the fifth follows from the first four. We prefer to keep all of them for symmetry reasons.

Our next goal is to show that the system of integrability conditions
\eqref{int} is in involution, and its general solution depends
on six arbitrary functions of one variable. \medskip

\begin{theorem} \label{six} 
The general 2D integrable Lagrangian
density $f$ depends on six arbitrary functions of one variable.
\end{theorem}

\medskip \centerline{\bf Proof:}

\medskip

Let us introduce the new dependent variables
\begin{equation*}
s=(s_1, s_2, s_3, s_4, s_5, s_6)^T=(f_{aa}, f_{ab}, f_{ac}, f_{bb}, f_{bc},
f_{cc})^T,
\end{equation*}
which satisfy the obvious consistency conditions such as
\begin{equation*}
\begin{array}{c}
(s_1)_b=(s_2)_a, \ (s_1)_c=(s_3)_a, \ (s_2)_b=(s_4)_a, \ (s_2)_c=(s_3)_b=(s_5)_a,\\
\ \\
(s_3)_c=(s_6)_a, \ (s_4)_c=(s_5)_b, \ (s_5)_c=(s_6)_b.
\end{array}
\end{equation*}
Taking these consistency conditions along with the four integrability
conditions \eqref{int} (also rewritten in terms of $s$
-variables) we obtain a system of twelve first-order quasilinear equations
for $s_i(a, b, c)$ which can be represented in the form of two six-component systems
of hydrodynamic type,
\begin{equation}  \label{s}
s_a={\bf P}(s)s_c, \quad s_b={\bf Q}(s)s_c,
\end{equation}
where $\bf{P, Q}$ are the following $6\times 6$ matrices:
\begin{equation*}
\mathbf{P}=\frac{1}{s_6}\left(
\begin{array}{cccccc}
{2s_{3}+s_{4}} & -2{s_{2}} & -{s_{1}} &
{s_{1}} & 0 & 0 \\
2{s_{5}} & 0 & -2{s_{2}} & 0 &{s_{1}} &
0 \\
s_6 & 0 & 0 & 0 & 0 & 0 \\
s_6 & 2{s_{5}} & -{2s_{3}-s_{4}}& 0 & 0 &{s_{1}
} \\
0 & s_6 & 0 & 0 & 0 & 0 \\
0 & 0 & s_6 & 0 & 0 & 0
\end{array}
\right),
\end{equation*}
\begin{equation*}
\mathbf{Q}=\frac{1}{s_6}\left(
\begin{array}{cccccc}
2{s_{5}} & 0 & -2{s_{2}} & 0 & {s_{1}} &
0 \\
s_6 & 2{s_{5}} & -{2s_{3}-s_{4}} & 0 & 0 & {s_{1}
} \\
0 & s_6 & 0 & 0 & 0 & 0 \\
0 & 2s_6 & -2{s_{5}} & 2{s_{5}} & -{2s_{3}-s_{4}} & 2{s_{2}} \\
0 & 0 & 0 & s_6 & 0 & 0 \\
0 & 0 & 0 & 0 & s_6 & 0
\end{array}
\right) .
\end{equation*}
Equations \eqref{s} possess six conserved densities $s_2,\ s_4,\ s_5,\ s_1s_5-s_2s_3,\ s_2s_6-s_3s_5, \ s_1s_6+s_2s_5-s_3s_4-s_3^2$ which satisfy the equations
\begin{equation}\label{cons}
\begin{array}{c}
(s_2)_a=(s_1)_b,\qquad (s_4)_a=(s_2)_b,\qquad (s_5)_a=(s_3)_b,\\
\ \\
(s_2)_c=(s_3)_b,\qquad (s_4)_c=(s_5)_b,\qquad (s_5)_c=(s_6)_b,\\
\ \\
2(s_1s_5-s_2s_3)_a=(s_1 s_4-s_2^2)_b,\qquad 2(s_2s_6-s_3s_5)_a=(s_1 s_6-s_3^2)_b,\\
\ \\
2(s_1s_5-s_2s_3)_c=(s_1 s_6-s_3^2)_b,\qquad 2(s_2s_6-s_3s_5)_c=(s_4 s_6-s_5^2)_b,\\
\ \\
(s_1s_6+s_2s_5-s_3s_4-s_3^2)_a=(s_1 s_5-s_2s_3)_b,\\
\ \\
(s_1s_6+s_2s_5-s_3s_4-s_3^2)_c=(s_2 s_6-s_3s_5)_b.
\end{array}
\end{equation}
Direct calculation shows that systems (\ref{s}) commute, that is, $
s_{ab}=s_{ba}$. Thus, equations (\ref{s}) are in involution, and their general common solution depends on six arbitrary
functions of one variable, namely, the Cauchy data $s_i(0, 0, c)$. This finishes the proof.

\medskip
 \noindent \textbf{Remark 2.} Relations \eqref{int} and \eqref{cons}
imply that there exists a potential $\rho $ such that
\begin{equation}\label{rho}
\begin{array}{c}
\rho _{aa}=f_{aa}f_{bb}-f_{ab}^{2},\quad \rho
_{ac}=f_{aa}f_{cc}-f_{ac}^{2},\quad \rho _{cc}=f_{cc}f_{bb}-f_{bc}^{2},\\
\ \\
\rho _{ab}=2\left( f_{bc}f_{aa}-f_{ac}f_{ab}\right) ,\quad \rho_{bc}=2\left( f_{ab}f_{cc}-f_{ac}f_{bc}\right), \\
\ \\
\rho _{bb}=2(f_{ab}f_{bc}-f_{ac}f_{bb}+f_{aa}f_{cc}-f_{ac}^{2}).
\end{array}
\end{equation}
\medskip

\medskip
 \noindent \textbf{Remark 3.}
 System (\ref{int}) 
 possesses a Lax pair
$$
\psi_a=\lambda {\bf K} \psi, \quad \psi_b=\lambda {\bf L} \psi, \quad \psi_c=\lambda {\bf M} \psi, 
$$
where $\lambda$ is a spectral parameter and the $4\times 4$ matrices ${\bf K}, {\bf L}, {\bf M}$ are defined as
\begin{equation*}
{\bf K}=\begin{pmatrix}
0 & f_{ac} & 0 & 1 \\
-f_{aa} & -f_{ab} & 0 & 0 \\
-\frac{1}{2}\rho _{ab} & -\rho _{ac} & -f_{ab} & f_{ac} \\
-\rho _{aa} & -\frac{1}{2}\rho _{ab} & -f_{aa} & 0
\end{pmatrix},
\quad
{\bf L}=\begin{pmatrix}
0 & f_{bc} & 1 & 0 \\
-f_{ab} & -f_{bb} & 0 & 1 \\
-\frac{1}{2}\rho _{bb} & -\rho _{bc} & -f_{bb} & f_{bc} \\
-\rho _{ab} & -\frac{1}{2}\rho _{bb} & -f_{ab} & 0
\end{pmatrix},
\end{equation*}
\begin{equation*}
{\bf M}=\begin{pmatrix}
0 & f_{cc} & 0 & 0 \\
-f_{ac} & -f_{bc} & 1 & 0 \\
-\frac{1}{2}\rho _{bc} & -\rho _{cc} & -f_{bc} & f_{cc} \\
-\rho _{ac} & -\frac{1}{2}\rho _{bc} & -f_{ac} & 0
\end{pmatrix}.
\end{equation*}

 \medskip

\noindent
\textbf{Remark 4.} We have verified that both systems (\ref
{s}) are linearly degenerate and non-diagonalisable (their Haantjes tensor
does not vanish). This suggests that integrable Lagrangian densities (\ref{2}) are related to the associativity (WDVV) equations where analogous
commuting six-component systems were obtained in \cite{FM}, see also \cite{PV1, PV2} for related results. Such a link indeed exists, and is discussed in Section \ref{sec:i0}.


\subsection{Equivalence group in 2D}
\label{sec:equiv}

Let $\mathbf{U}$ be the $2\times 2$  Hessian matrix of the function $u(x,y)$. Integrable Lagrangians of type (\ref{2})  are invariant under ${\rm Sp}(4, \mathbb{R})$-symmetry
\begin{equation}
\mathbf{U}\rightarrow (\mathbf{AU}+\mathbf{B})(\mathbf{CU}+\mathbf{D})^{-1}, \quad f\to \frac{f}{\det (\mathbf{CU}+\mathbf{D})},
\label{sp4}
\end{equation}
where the
matrix
\begin{equation*}
\left(
\begin{array}{cc}
\mathbf{A} & \mathbf{B} \\
\mathbf{C} & \mathbf{D}
\end{array}
\right)
\end{equation*}
belongs to the symplectic group ${\rm Sp}(4, \mathbb{R})$ (here $\mathbf{A},\mathbf{B},\mathbf{
C},\mathbf{D}$ are $2\times 2$ matrices). Symmetry (\ref{sp4}) suggests a relation to  Siegel modular forms (the density $f$ transforms as a genus two Siegel modular form of weight $-1$). This symmetry corresponds to
linear symplectic transformations of the four-dimensional jet space with
coordinates $u_{x},u_{y}, x, y$. Furthermore, integrable Lagrangians (\ref{2})
are invariant under rescalings of $f$, as well as under the addition of a
`null-Lagrangian', namely, transformations of the form
\begin{equation}
f\rightarrow \lambda_0 f+\lambda_1(u_{xx}u_{yy}-u_{xy}^{2})+\lambda_2u_{xx}+\lambda_3u_{xy}+\lambda_4u_{yy}+\lambda_5,
\label{null2}
\end{equation}
which do not effect the Euler-Lagrange equations. Transformations (\ref{sp4}
) and (\ref{null2}) generate a group of dimension $10+6=16$ which preserves
the class of integrable Lagrangians (\ref{2}). These equivalence
transformations will be utilised to simplify the classification results in
Section \ref{sec:ex}. For instance, modulo equivalence transformations the Lagrangian density (\ref{Mir}) is equivalent to $f=u_{xy}(u_{xx}^2-u_{yy}^2)$.

\subsection{Integrability conditions via generalised Rankin-Cohen brackets}
\label{sec:RC}

Integrability conditions (\ref{int}) possess a compact formulation via  higher genus Rankin-Cohen
(Eholzer-Ibukiyama) brackets for Siegel modular forms  \cite{Ibu}. This does not come as something unexpected, indeed, the integrability conditions possess ${\rm Sp}(4)$-invariance (\ref{sp4}) and, therefore, should be expressible via ${\rm Sp}(4)$-invariant operations. Here we mainly follow  \cite{Miya, Ibukiyama1}, which specialised the general results of \cite{Ibu} to the genus two case. 
Let us introduce two matrix differential operators
$$
R=\left(\begin{array}{cc}
\partial_a&\frac{1}{2}\partial_b\\
\frac{1}{2}\partial_b& \partial_c
\end{array}\right), \quad 
S=\left(\begin{array}{cc}
\partial_{\tilde a}&\frac{1}{2}\partial_{\tilde b}\\
\frac{1}{2}\partial_{\tilde b}& \partial_{\tilde c}
\end{array}\right),
$$
and define the operators $P_0, P_1, P_2$ via the expansion
$$
\det (R+\lambda S)=P_0+\lambda P_1+\lambda^2 P_2.
$$
Explicitly, we have
$$
P_0=\partial_a\partial_c-\frac{1}{4}\partial_b^2, \quad 
P_1=\partial_a\partial_{\tilde c}+\partial_c\partial_{\tilde a}-\frac{1}{2}\partial_b\partial_{\tilde b}, \quad
P_2=\partial_{\tilde a}\partial_{\tilde c}-\frac{1}{4}\partial_{\tilde b}^2.
$$
Let us also define two operators $Y_1$, $Y_2$ depending on the auxiliary parameters $\xi=(\xi_1, \xi_2)$ by the formulae
$$
Y_1=\xi R\xi^T=\xi_1^2\partial_a+\xi_1\xi_2\partial_b+\xi_2^2\partial_c, \quad
 Y_2=\xi S\xi^T=\xi_1^2\partial_{\tilde a}+\xi_1\xi_2\partial_{\tilde b}+\xi_2^2\partial_{\tilde c}.
$$
Finally, we introduce the $\xi$-dependent operator 
$$
v=(\partial_a\partial_{\tilde b}-\partial_b\partial_{\tilde a})\xi_1^2+2(\partial_a\partial_{\tilde c}-\partial_c\partial_{\tilde a})\xi_1\xi_2+(\partial_b\partial_{\tilde c}-\partial_c\partial_{\tilde b})\xi_2^2.
$$
Then integrability conditions (\ref{int}) can be represented in the Hirota-type bilinear form
\begin{equation}\label{RC}
(P_1Y_1v-2P_0Y_2v)[f(a, b, c)\cdot f(\tilde a, \tilde b, \tilde c)]\biggr \rvert_
{\tilde a=a, \ \tilde b=b,\ \tilde c=c}=0.
\end{equation}
Here the left-hand side is a homogeneous quartic  in $\xi_1, \xi_2$, with five nontrivial components. Equating them to zero we obtain all of the  five integrability conditions (\ref{int}).

\medskip

\noindent {\bf Remark 5.} It follows from \cite{Miya}, Proposition 2.3,   that if $f$ transforms as in 
(\ref{sp4}), that is, as a weight $-1$ Siegel modular form, then the left-hand side of (\ref{RC}) transforms as a vector-valued Siegel modular form with values in the representation $ {\rm Sym}_4\otimes \det$ of ${\rm GL}(2, \mathbb{C})$. 

\medskip

\noindent {\bf Remark 6.} The principal symbol of the Euler-Lagrange equation (\ref{Euler})  is given by a compact formula in terms of the operator $Y_1$:
$$
Y_1^2[f]=f_{aa}\xi_1^4+2f_{ab}\xi_1^3\xi_2+(2f_{ac}+f_{bb})\xi_1^2\xi_2^2+2f_{bc}\xi_1\xi_2^3+f_{cc}\xi_2^4.
$$
This expression transforms as a vector-valued Siegel modular form with values in the representation $ {\rm Sym}_4\otimes \det^{-1}$. 

\subsection{Integrable Lagrangians and classical Newtonian equations}
\label{sec:1/2}

Here we sketch the derivation of the Lagrangian density (\ref{dKP}). Consider a classical Newtonian equation
\begin{equation*}
\ddot{x}=-U_{x}
\end{equation*}
where $U(x, t)$ is the potential function,  $x=x(t)$,  and dot denotes differentiation by $t$. This equation can be written in the
canonical Hamiltonian form
\begin{equation*}
\dot{x}=p,\text{ \ }\dot{p}=-U_{x}. 
\end{equation*}
To be Liouville integrable, this Hamiltonian system should be equipped with an extra
first integral $F(t,x, p)$ such that
\begin{equation}\label{xp1}
\frac{dF}{dt}\equiv \frac{\partial F}{\partial t}+ \frac{\partial F}{\partial x}\dot x+\frac{\partial F}{\partial p}\dot{p}=F_t+pF_x-U_xF_p=0.
\end{equation}
First integrals $F$ polynomial in the momentum $p$ were thoroughly
investigated in \cite{Drach}, and later in \cite{Kozlov, PT}. In particular, the following cases have been studied:
\begin{equation*}
F=\frac{p^{3}}{3}+Up+V,\quad F=\frac{p^{4}}{4}+Up^{2}+Vp+W,
\quad F=\frac{p^{5}}{5}+Up^{3}+Vp^{2}+Wp+Q.
\end{equation*}
In the last (fifth-order) case, equation  (\ref{xp1}) implies the following quasilinear system for the coefficients:
\begin{equation*}
U_{t}+V_{x}=0,\text{ \ }V_{t}+W_{x}=3UU_{x},\text{ \ }W_{t}+Q_{x}=2VU_{x},
\text{ }Q_{t}=WU_{x}.
\end{equation*}
Let us introduce a  potential  $u$ such that 
$U=u_{xx}$, $V=-u _{xt}$, $W=\frac{3}{2}u _{xx}^{2}+u_{tt}$. Then the first two equations will be satisfied identically, while the last two  imply
\begin{equation*}
Q_{x}=-2u_{xt}u_{xxx}-3u_{xx}u_{xxt}-u_{ttt}, \quad
Q_{t}=\frac{3}{2}u_{xx}^{2}u_{xxx}+u_{tt}u_{xxx}.
\end{equation*}
The compatibility condition of these equations for $Q$ leads to a fourth-order PDE for $u$,
\begin{equation*}
u_{tttt}+\frac{3}{2}u_{xx}^{2}u_{xxxx}+3u_{xx}u_{xxx}^{2}+u_{tt}u_{xxxx}+2u_{xt}u_{xxxt}+3u_{xx}u_{xxtt}+3u_{xtt}u_{xxx}+3u_{xxt}^{2}=0,
\end{equation*}
which is nothing but the Euler-Lagrange equation for the second-order Lagrangian
\begin{equation*}
S=\int \left[u_{tt}^{2}+u_{xx}^{2}u_{tt}+u_{xx}u_{xt}^{2}+\frac{1}{4}u_{xx}^{4}\right] dxdt,
\end{equation*}
whose density is identical to (\ref{dKP}) up to relabelling $t\leftrightarrow y$.

\subsection{Integrable Lagrangians and WDVV  equations}
\label{sec:i0}

Let $F(t_{1},\dots, t_{n})$ be a function  of $n$ independent variables such that the symmetric matrix
$$
\eta_{ij}=\partial_1\partial_i\partial_jF
$$
is constant and non-degenerate (thus, $t_1$ is a marked variable), and the coefficients
$$
c^i_{jk}=\eta^{is}\partial_s\partial_j\partial_kF
$$
satisfy the associativity condition $c^s_{ij}c^p_{sk}=c^s_{kj}c^p_{si}$; here $i, j, k \in \{1, \dots, n\}$. These  requirements  impose a nonlinear system of  third-order PDEs for the prepotential $F$, the so-called associativity (WDVV) equations which were discovered in the beginning of 1990s by Witten, Dijkgraaf, Verlinde and Verlinde in the context of two-dimensional topological field theory.
Geometry and integrability of WDVV equations has been thoroughly studied by Dubrovin, culminating in the remarkable theory of Frobenius manifolds \cite{Dub}. An important ingredient of this theory is an integrable hydrodynamic hierarchy whose `primary' part is defined by $n-1$  commuting Hamiltonian flows
\begin{equation}
\label{prim}
\partial_{T_{\alpha}}t_i=c^i_{\alpha  k}\partial_Xt_k=\partial_X(\eta^{is}\partial_s\partial_{\alpha}F)
\end{equation}
where  $T_{\alpha}$ are the higher `times'; here $T_1=X$. The flows (\ref{prim}) are manifestly Hamiltonian with the Hamiltonian operator $\eta^{is}\frac{d}{dX}$ and the Hamiltonian density $\partial_{\alpha} F$. Note that WDVV equations are equivalent to the requirement of commutativity of these flows.

\medskip

We will need a particular case of the general construction when $n=4$ and the matrix $\eta$ is anti-diagonal, which corresponds to prepotentials
\begin{equation}\label{prep}
F(t_{1},t_{2},t_{3},t_{4})=\frac{1}{2}
t_{1}^{2}t_{4}+t_{1}t_{2}t_{3}+W(t_{2},t_{3},t_{4}).
\end{equation}
The corresponding primary flows (\ref{prim}) take the form
\begin{equation}\label{prim1}
\begin{array}{c}
\partial_{T_2}t_1=\partial_X(\partial_4\partial_2F), \quad \partial_{T_2}t_2=\partial_X(\partial_3\partial_2F), \quad \partial_{T_2}t_3=\partial_X(\partial_2\partial_2F), \quad \partial_{T_2}t_4=\partial_X(\partial_1\partial_2F),\\
\ \\
\partial_{T_3}t_1=\partial_X(\partial_4\partial_3F), \quad \partial_{T_3}t_2=\partial_X(\partial_3\partial_3F), \quad \partial_{T_3}t_3=\partial_X(\partial_2\partial_3F), \quad \partial_{T_3}t_4=\partial_X(\partial_1\partial_3F),\\
\ \\
\partial_{T_4}t_1=\partial_X(\partial_4\partial_4F), \quad \partial_{T_4}t_2=\partial_X(\partial_3\partial_4F), \quad \partial_{T_4}t_3=\partial_X(\partial_2\partial_4F), \quad \partial_{T_4}t_4=\partial_X(\partial_1\partial_4F),\\
\end{array}
\end{equation}
which are Hamiltonian systems with the Hamiltonian  densities
$$
\partial_2F=t_1t_3+\partial_2W, \quad \partial_3F=t_1t_2+\partial_3W, \quad \partial_4F=\frac{1}{2}t_1^2+\partial_4W,
$$
respectively. In compact form, equations (\ref{prim1}) can be represented as
$$
\partial_{T_{\alpha}}t_i=\partial_X(\partial_{5-i}\partial_{\alpha}F), \quad i =1, 2, 3, 4,\quad \alpha=2,3,4.
$$
Setting $(t_1, t_2, t_3, t_4)=(P, B, C, A)$ we obtain
\begin{equation*}
F=\frac{1}{2}P^2A+PBC+W(B,C, A).
\end{equation*}
In this case WDVV equations reduce to the following system of four PDEs for $W$:
\begin{align}\label{W1}
\begin{split}
W_{AAA}&=W_{ABC}^{2}+W_{ABB}W_{ACC}-W_{AAB}W_{BCC}-W_{AAC}W_{BBC},\\
W_{AAB}&=W_{BBB}W_{ACC}-W_{ABB}W_{BCC}, \\
W_{AAC}&=W_{ABB}W_{CCC}-W_{ACC}W_{BBC}, \\
2W_{ABC} &=W_{BBB}W_{CCC}-W_{BBC}W_{BCC}.
\end{split}
\end{align}
The corresponding primary flows (\ref{prim1}) take the form
\begin{equation}\label{prim2}
\begin{array}{c}
A_{T_2}=C _{X},\text{ \ }B_{T_2}=\left(P+ W_{BC}\right) _{X},
\text{ \ }C_{T_2}=\left( W_{BB}\right) _{X},\text{ \ }P_{T_2}=\left(W_{AB}\right)
_{X},  \\
\ \\
A_{T_3}=B _{X},\text{ \ }B_{T_3}=\left(W_{CC}\right) _{X},
\text{ \ }C_{T_3}=\left(P+W_{BC}\right) _{X},\text{ \ }P_{T_3}=\left(W_{AC}\right)
_{X},  \\
\ \\
A_{T_4}=P_{X},\text{ \ }B_{T_4}=\left(W_{AC}\right) _{X},
\text{ \ }C_{T_4}=\left(W_{AB}\right) _{X},\text{ \ }P_{T_4}=\left(W_{AA}\right)
_{X}.
\end{array}
\end{equation}
Note that  system (\ref{prim2})$_2$ coincides with (\ref{H}) under the identification $h=W_C$, thus establishing a link between WDVV equations and integrable Lagrangians. This link can be summarised as follows:

\begin{itemize}
\item Take  prepotential  of type (\ref{prep}), set $(t_2, t_3, t_4)=(B, C, A)$ and define $h(A, B, C)=W_C$.

\item Reconstruct Lagrangian density $f(a, b, c)$ by applying partial Legendre transform to $h(A, B, C)$:
$$
a=A,\ b=B,\ c=h_C, \ f=Ch_C-h, \ f_a=-h_A, \ f_b=-h_B, \ f_{c}=C.
$$
\end{itemize}
Examples of  calculations of this kind will be given in Section \ref{sec:WDVVex}.

 \medskip

\noindent
\textbf{Remark 7.} 
Conversely, given a Lagrangian density $f(a,b,c)$, the 
corresponding prepotential $W(A, B, C)$ can be reconstructed by the formulae
\begin{equation*}
W_{AA}=-\rho _{a},\text{ \ }W_{AB}=-\frac{1}{2}\rho _{b},\text{ \ }
W_{AC}=-f_{a},
\end{equation*}
\begin{equation*}
W_{BB}=-\rho _{c},\text{ \ }W_{BC}=-f_{b},\text{ \ }W_{CC}=c.
\end{equation*}
\begin{equation*}
A=a,\text{ \ }B=b,\text{ \ }C=f_{c},
\end{equation*}
where $\rho$ is defined by formulae (\ref{rho}), see Section \ref{sec:LWDVV}.

\subsection{Examples of integrable Lagrangians in 2D}

\label{sec:ex}

In this section we present explicit examples of integrable Lagrangian
densities $f$ obtained by assuming a suitable ansatz for $f$ and computing
the corresponding integrability conditions \eqref{int}. This gives a whole range of integrable densities expressible via polynomials, elementary functions, Jacobi theta functions and dilogarithms. 

\subsubsection{Integrable Lagrangian densities of the form $f=g(u_{xx},
u_{yy})$}

In this case the integrability conditions lead to the only constraint $
g_{aa}g_{cc}-g_{ac}^2=k$ where $k=const$. Its solutions can be represented
parametrically, thus, for $k=0$ (parabolic case) and $k=-1$ (hyperbolic
case) we obtain the general solution in parametric form:
\begin{equation*}
a=p^{\prime }(w)v+q^{\prime }(w),\quad c=v,\quad f= w[p^{\prime
}(w)v+q^{\prime }(w)]-[p(w)v+q(w)],
\end{equation*}
and
\begin{equation*}
a=p^{\prime }(w+v)+q^{\prime }(w-v),\quad c=v,\quad f=w [p^{\prime
}(w+v)+q^{\prime }(w-v)]-[p(w+v)+q(w-v)],
\end{equation*}
respectively; here $p$ and $q$ are arbitrary functions and prime denotes differentiation.

\subsubsection{Integrable Lagrangian densities of the form $
f=e^{u_{xx}}g(u_{xy}, u_{yy})$}
\label{sec:Jac}

We will show that the generic integrable density of this form corresponds to
$$
g(b, c)=[\Delta(ic/\pi)]^{-1/8}\theta_1(b, ic/\pi)
$$
where $\Delta$ is the modular discriminant and $\theta_1$ is the Jacobi theta function.
The details are as follows. Substituting $f=e^a g(b,c)$ into the integrability conditions \eqref{int} one obtains
\begin{equation}  \label{5_1}
gg_{bcc}=3g_{cc}g_{b}-2 g_{bc} g_{c},
\end{equation}
\begin{equation}  \label{5_2}
gg_{bbb}=g_{b}g_{bb}+4g_{bc}g-4g_{b}g_{c},
\end{equation}
\begin{equation}  \label{5_3}
gg_{ccc}=g_{c}g_{cc} +2 g_{cc} g_{bb}-2 (g_{bc})^2,
\end{equation}
\begin{equation}  \label{5_4}
gg_{bbc}=2g_{b}g_{bc}-g_{c}g_{bb}+2 g g_{cc}-2(g_{c})^2.
\end{equation}
This over-determined system for $g$ is in involution, and can be solved as
follows. First of all, equation \eqref{5_2} implies
\begin{equation*}
\left(\frac{g_{bb}}{g}\right)_b=\left(\frac{4 g_{c}}{g}\right)_b,
\end{equation*}
so that one can set
\begin{equation}  \label{5_gc}
g_{c}=\frac{1}{4}(g_{bb}-hg)
\end{equation}
where $h$ is a function of $c$ only. Using (\ref{5_gc}), both \eqref{5_1}
and \eqref{5_4} reduce to
\begin{equation}  \label{5_gbbbb}
g_{bbbb}g-4g_{b}g_{bbb}+3 g_{bb}^2=4h(gg_{bb}-g_{b}^2)-4h^{\prime }g^2;
\end{equation}
here prime denotes differentiation by $c$. Modulo \eqref{5_gc} and \eqref{5_gbbbb}, equation \eqref{5_3} implies
\begin{equation}  \label{5_gbbb2}
g_{bbb}^2g^2+g_{bbb}(4 g_{b}^3-6gg_{b}g_{bb})-3g_{b}^2 g_{bb}^2+4 gg_{bb}^3
=4h(g_{b}^2-gg_{bb})^2+8 h^{\prime}g^2(g_{b}^2-gg_{bb})+\frac{8}{3}h^{\prime
\prime }g^4.
\end{equation}
Note that \eqref{5_gbbb2} can be obtained from \eqref{5_gbbbb} by
differentiating it with respect to $c$, and using (\ref{5_gc}), \eqref{5_gbbbb}. Similarly,
differentiating \eqref{5_gbbb2} with respect to $c$ we obtain the Chazy
equation \cite{Chazy} for $h$:
\begin{equation}  \label{chazy}
h^{\prime \prime \prime }=2hh^{\prime \prime }-3h^{\prime 2}.
\end{equation}
Equations (\ref{5_gbbbb}) and (\ref{5_gbbb2}) can be simplified by the
substitution $v=-(\ln g)_{bb}$, which implies
\begin{equation}  \label{wp1}
v_{bb}=6v^2+4hv+4h^{\prime }
\end{equation}
and
\begin{equation}  \label{wp2}
v_b^2=4v^3+4hv^2+8h^{\prime }v+\frac{8}{3}h^{\prime \prime },
\end{equation}
respectively. Since (\ref{wp1}) follows from (\ref{wp2}) via differentiating
with respect to $b$, we end up with the following compact form of
integrability conditions (\ref{5_1})-(\ref{5_4}):
\begin{equation}  \label{comp}
g_{c}=\frac{1}{4}(g_{bb}-hg), \quad v=-(\ln g)_{bb}, \quad
v_b^2=4v^3+4hv^2+8h^{\prime }v+\frac{8}{3}h^{\prime \prime };
\end{equation}
here $h$ solves the Chazy equation (\ref{chazy}). We recall that modulo the
natural ${\rm SL}(2, \mathbb{R})$-symmetry \cite{CO}, the Chazy equation possesses three
non-equivalent solutions: $h=0, \ h=1$ and $h=\frac{1}{2}\frac{\Delta'}{\Delta}$ where $\Delta$ is the
modular discriminant. These three solutions (which correspond to rational,
trigonometric and elliptic cases of the Weierstrass $\wp$-function equation
in (\ref{comp})) are considered separately below. Note that both the rational and trigonometric cases lead to degenerate Lagrangians, so only the elliptic case is of interest.

\medskip

\noindent\textbf{Rational case} $h=0$. In this case equations (\ref{comp})
simplify to
\begin{equation*}
g_{c}=\frac{1}{4}g_{bb}, \quad v=-(\ln g)_{bb}, \quad v_b^2=4v^3,
\end{equation*}
which are straightforward to solve. Modulo unessential constants the generic
solution of these equations is $g=e^{2\mu b+\mu^2c}(b+\mu c)$ where $
\mu=const$. The corresponding Lagrangian density $f$ takes the form
\begin{equation*}
f=e^{u_{xx}+2\mu u_{xy}+\mu^2u_{yy}}(u_{xy}+\mu u_{yy}).
\end{equation*}
Note that the change of independent variables $x=\tilde x, \ y=\tilde y+\mu
\tilde x$ brings this Lagrangian to the degenerate form $\tilde
f=e^{u_{\tilde x\tilde x}}u_{\tilde x\tilde y}$ (the order of the
corresponding Euler-Lagrange equation can be reduced by two by setting $
v=u_{\tilde x}$).

\medskip

\noindent\textbf{Trigonometric case} $h=1$. In this case equations (\ref
{comp}) simplify to
\begin{equation*}
g_{c}=\frac{1}{4}(g_{bb}-g), \quad v=-(\ln g)_{bb}, \quad
v_b^2=4v^3+4v^2,
\end{equation*}
which are also straightforward to solve. Modulo unessential constants the
generic solution of these equations is $g=e^{2\mu b+\mu^2c}\sinh(b+\mu c)$
where $\mu=const$. The corresponding Lagrangian density $f$ takes the form
\begin{equation*}
f=e^{u_{xx}+2\mu u_{xy}+\mu^2u_{yy}}\sinh (u_{xy}+\mu u_{yy}).
\end{equation*}
Note that the same change of variables as in the rational case brings this
Lagrangian to the degenerate form $\tilde f=e^{u_{\tilde x\tilde x}}\sinh
u_{\tilde x\tilde y}$.

\medskip

\noindent\textbf{Elliptic case} $h=\frac{1}{2}\frac{\Delta'}{\Delta}$, see e.g. \cite{Takh}. Here  the modular discriminant $\Delta$ is given by the formula 
$$
\Delta(c)=(2\pi)^{12}q\prod_{1}^{\infty}(1-q^n)^{24}, \quad q=e^{2\pi i c},
$$
recall that $h$ has the $q$-expansion
$$
h(c)=\pi i E_2=\pi i\left(1-24\sum_{n=1}^{\infty}\sigma_1(n)q^n\right)
$$
where $E_2$ is the Eisenstein series (here $\sigma_1(n)$ is the  divisor function). Setting $g(b, c)=[\Delta(c)]^{-1/8} w(b, c)$ we see that the first equation (\ref{comp})  becomes the heat equation for $w$: 
\begin{equation}  \label{comp1}
w_{c}=\frac{1}{4}w_{bb}, \quad v=-(\ln w)_{bb}, \quad
v_b^2=4v^3+4hv^2+8h^{\prime }v+\frac{8}{3}h^{\prime \prime }.
\end{equation}
The general solution of system (\ref{comp1}) was constructed in \cite{BLP}:
$$
w(b, c)=\Delta^{1/8}\sigma(b, g_2, g_3)e^{b^2h/6}
$$
where $\sigma$ is the Weierstrass sigma function with the invariants $g_{2}=\frac{4}{3}h^{2}-8h^{\prime },\text{ \ }g_{3}=-\frac{8}{27}h^{3}+\frac{8}{3}hh^{\prime }-\frac{8}{3}h^{\prime \prime }$.
Note that  $\Delta=\pi^6(g_2^3-27g_3^2)$. 
 Thus,
$$
g(b, c)=\sigma(b, g_2, g_3)e^{b^2h/6}.
$$
\noindent{\bf Remark 8.} An alternative (real-valued) representation of the general solution of system (\ref{comp1})  in terms of the Jacobi theta function $\theta_1$ is as follows:
$$
w(b, c)=\theta_1(b, ic/\pi)=2\sum_{n=0}^{\infty} (-1)^ne^{-(n+1/2)^2c}\sin[(2n+1)b];
$$
here for $h$ in the last equation (\ref{comp1}) one has to use
$$
\frac{i}{\pi}h(ic/\pi)=-1+24\sum_{n=1}^{\infty}\sigma_1(n)e^{-2nc}=-1+24(e^{-2c}+3e^{-4c}+4e^{-6c}+7e^{-8c}+\dots),
$$
which is another (real-valued) solution of the Chazy equation (note that the Chazy equation is invariant under the scaling symmetry $h(c)\to \lambda h(\lambda c)$. Thus,
$$
g(b, c)=[\Delta(ic/\pi)]^{-1/8}\theta_1(b, ic/\pi).
$$
 Note that the function   $ \triangle^{-1/8}(\tau)\theta_1(z, \tau)$ appears in the theory of weak Jacobi forms (it is a holomorphic weak Jacobi form of weight $-1$ and index $1/2$).  

\subsubsection{Integrable Lagrangian densities polynomial in $e^{u_{xx}}$ and $e^{u_{yy}}$}
\label{sec:di}

Here we describe integrable Lagrangian densities $f$ that are linear/quadratic  in $e^{u_{xx}}$ and $e^{u_{yy}}$, the coefficients being functions of ${u_{xy}}$ only.
\medskip

\noindent {\bf Linear case:}
$$
f=p_0+p_1 e^a + p_2 e^c.
$$
Substituting this ansatz into the integrability conditions (and assuming $p_1, p_2$ to be nonzero) we obtain a system of ODEs for the coefficients $p_i(b)$ which, modulo  equivalence transformations, can be simplified to  
$$
p_1=p_2=p, \quad p''=p, \quad p_0''=\alpha/p;
$$
here $\alpha=const$ (which can be set equal to $1$ if nonzero) and prime denotes differentiation by $b$. Modulo equivalence transformations, these equations possess two essentially different solutions:
$$
f=\alpha e^{-b}+(e^a+e^c)e^b \qquad {\rm and} \qquad f=\alpha q(b)+(e^a+e^c)\sinh b,
$$
where  the function $q(b)$ satisfies $q''=\frac{1}{\sinh b}$. This implies $q'=\ln\frac{ 1-e^b}{1+e^b}$, and another integration gives
$$
q(b)=Li_2(-e^b)-Li_2(e^b)
$$
where $Li_2$ is the dilogarithm function: $(Li_2(x))'=-\frac{\ln (1-x)}{x}$.

\medskip

\noindent {\bf Quadratic case:}
$$
f=p_0+p_1 e^a + p_2 e^c +p_3 e^{2a}+p_4 e^{a+c}+p_5 e^{2c}.
$$
Substituting this ansatz into the integrability conditions we obtain a large system of ODEs for the coefficients $p_i(b)$ which,  modulo equivalence transformations,  lead to
the following integrable  densities (here we only present those examples that are not reducible to the  linear case by a change of variables):
$$
f= e^{kb+a+c},
\qquad
f=  e^{\frac{4}{\sqrt  3}b+a+c}+ e^{\frac{2}{\sqrt 3}b+2c},
\qquad
f=\alpha e^{-\frac{1}{\sqrt2}b}+\alpha e^{ \frac{1}{\sqrt2}b+a }+e^{\frac{{1}}{\sqrt 2}b+c}+e^{\frac{3}{\sqrt 2}b+a+c},
$$
$$
f= p e^{2a}+2p^2 e^{a+c} +p e^{2c},\quad p=\cosh \left( \frac{2}{\sqrt 3}b\right).
$$






\subsubsection{Integrable Lagrangian densities from WDVV prepotentials}
\label{sec:WDVVex}

In this section we discuss  polynomial prepotentials $F$ of type (\ref{prep}) associated with finite Coxeter groups $W$ as given in  \cite{Dub1}, p. 107. Applying the procedure outlined at the end of Section \ref{sec:i0} we compute the corresponding integrable Lagrangian densities $f$ which, in general, will be algebraic functions of $a, b, c$ (presented below up to appropriate scaling factors).

\medskip

\noindent {\bf Group $W(A_{4})$:}
\begin{equation*}
F={\frac{1}{2}}t_{1}^{2}\,t_{4}+t_{1}\,t_{2}\,t_{3}+{\frac{1}{2}}t_{2}^{3}+{
\frac{1}{3}}
t_{3}^{4}+6t_{2}t_{3}^{2}t_{4}+9t_{2}^{2}t_{4}^{2}+24t_{3}^{2}t_{4}^{3}+{
\frac{216}{5}}t_{4}^{6};
\end{equation*}
\begin{equation*}
f=\left(c- 48\,{a}^{3}-12\,ab \right)^{3/2}.
\end{equation*}
\noindent Swapping $t_2$ and $t_3$ (which is an obvious symmetry of WDVV equations) and following the same procedure gives a polynomial density $f$:
\begin{equation*}
F={\frac{1}{2}}t_{1}^{2}\,t_{4}+t_{1}\,t_{2}\,t_{3}+{\frac{1}{2}}t_{3}^{3}+{
\frac{1}{3}}
t_{2}^{4}+6t_{3}t_{2}^{2}t_{4}+9t_{3}^{2}t_{4}^{2}+24t_{2}^{2}t_{4}^{3}+{
\frac{216}{5}}t_{4}^{6};
\end{equation*}
\begin{equation*}
f=54\,{a}^{4}-6\,{a}^{2}c+\frac{1}{6}\,{c}^{2}-6\,{b}^{2}a.
\end{equation*}
\medskip {\bf Group $W(B_{4})$:}
\begin{equation*}
F={\frac{1}{2}}t_{1}^{2}\,t_{4}+t_{1}\,t_{2}\,t_{3}+{{{t_{2}}}^{3}}+{\frac{{{
t_{2}}\,{{{t_{3}}}^{3}}}}{3}}+3\,{{{t_{2}}}^{2}}\,{t_{3}}\,{t_{4}}+{\frac{{{{
{t_{3}}}^{4}}\,{t_{4}}}}{4}}
+3\,{t_{2}}\,{{{t_{3}}}^{2}}\,{{{t_{4}}}^{2}}+6\,{{{t_{2}}}^{2}}\,{{{t_{4}}}
^{3}}+{{{t_{3}}}^{3}}\,{{{t_{4}}}^{3}}+{\frac{{18\,{{{t_{3}}}^{2}}\,{{{t_{4}}
}^{5}}}}{5}}+{\frac{{18\,{{{t_{4}}}^{9}}}}{7}};
\end{equation*}
\begin{equation*}
f=2\,a{C}^{3}+ \left( 3\,{a}^{3}+b \right) {C}^{2}-3\,a{b}^{2},
\end{equation*}
where $C$ is defined by the quadratic   equation
$3 a C^2+(6 a^3+2 b) C +\frac{6}{5} a^2(6a^3+5b)=c.$
Swapping $t_2$ and $t_3$ gives a polynomial density $f$:
\begin{equation*}
F={\frac{1}{2}}t_{1}^{2}\,t_{4}+t_{1}\,t_{2}\,t_{3}+{{{t_{3}^{3}}}}+{\frac{{{
t_{2}^{3}}\,{{{t_{3}}}}}}{3}}+3\,{{{t_{2}}}}\,{t_{3}^{2}}\,{t_{4}}+{\frac{{{{
{t_{2}^{4}}}}\,{t_{4}}}}{4}}
+3\,{t_{2}^{2}}\,{{{t_{3}}}}\,{{{t_{4}^{2}}}}+6\,{{{t_{3}^{2}}}}\,{{{
t_{4}^{3}}}}+{{{t_{2}^{3}}}}\,{{{t_{4}^{3}}}}+{\frac{{18\,{{{t_{2}^{2}}}}\,{{
{t_{4}^{5}}}}}}{5}}+{\frac{{18\,{{{t_{4}^{9}}}}}}{7}};
\end{equation*}
\begin{equation*}
f=12\,{a}^{6}+12\,{a}^{4}b-2\,{a}^{3}c-bac+\frac{1}{12}\,{c}^{2}-\frac{1}{3}\,{b}^{3}.
\end{equation*}
\medskip {\bf Group $W(D_{4})$:}
\begin{equation*}
F={\frac{1}{2}}t_{1}^{2}\,t_{4}+t_{1}\,t_{2}
\,t_{3}+t_{2}^{3}t_{4}+t_{3}^{3}t_{4}+6t_{2}t_{3}t_{4}^{3}+{\frac{54}{35}}
t_{4}^{7};
\end{equation*}
\begin{equation*}
f={\frac {{c}^{2}}{12 a}}-6\,b{a}^{3}.
\end{equation*}
\medskip {\bf Group $W(F_{4})$:}
\begin{equation*}
F={\frac{1}{2}}t_{1}^{2}\,t_{4}+t_{1}\,t_{2}\,t_{3}+{\frac{t_{2}^{3}\,t_{4}}{
18}}+{\frac{3\,t_{3}^{4}\,t_{4}}{4}}+{\frac{t_{2}\,t_{3}^{2}\,t_{4}^{3}}{2}}
+{\frac{t_{2}^{2}\,t_{4}^{5}}{60}}+{\frac{t_{3}^{2}\,t_{4}^{7}}{28}}+{\frac{
t_{4}^{13}}{2^{4}\cdot 3^{2}\cdot 11\cdot 13}};
\end{equation*}
\begin{equation*}
f=\frac{1}{\sqrt a}  \left( {a}^{7}+14\,b{a}^{3}-14\,c \right) ^{3/2}.
\end{equation*}
Swapping $t_2$ and $t_3$ gives a rational density $f$:
\begin{equation*}
F={\frac{1}{2}}t_{1}^{2}\,t_{4}+t_{1}\,t_{2}\,t_{3}+{\frac{t_{3}^{3}\,t_{4}}{
18}}+{\frac{3\,t_{2}^{4}\,t_{4}}{4}}+{\frac{t_{2}^{2}\,t_{3}\,t_{4}^{3}}{2}}
+{\frac{t_{3}^{2}\,t_{4}^{5}}{60}}+{\frac{t_{2}^{2}\,t_{4}^{7}}{28}}+{\frac{
t_{4}^{13}}{2^{4}\cdot 3^{2}\cdot 11\cdot 13}};
\end{equation*}
\begin{equation*}
f={\frac {{a}^{9}}{600}}-\frac{1}{10}\,c{a}^{4}-\frac{1}{2}\,{b}^{2}{a}^{3}
+{\frac{3\,{c}^{2}}{2\,a}}.
\end{equation*}
\medskip {\bf Group $W(H_{4})$:}
\begin{equation*}
F={\frac{1}{2}}t_{1}^{2}\,t_{4}+t_{1}\,t_{2}\,t_{3}+{\frac{{{2\,{{t_{2}}}^{3}
}\,{t_{4}}}}{{3}}}+{\frac{{{{{t_{3}}}^{5}}\,{t_{4}}}}{{240}}}+{\frac{{{t_{2}}
\,{{{t_{3}}}^{3}}\,{{{t_{4}}}^{3}}}}{{18}}}+{\frac{{{{{t_{2}}}^{2}}\,{t_{3}}
\,{{{t_{4}}}^{5}}}}{{15}}}+{\frac{{{{{t_{3}}}^{4}}\,{{{t_{4}}}^{7}}}}{{
2^{3}\cdot 3^{3}\cdot 5}}}
\end{equation*}
\begin{equation*}
+{\frac{{{t_{2}}\,{{{t_{3}}}^{2}}\,{{{t_{4}}}^{9}}}}{{2\cdot 3^{4}\cdot 5}}}+
{\frac{{8\,{{{t_{2}}}^{2}}\,{{{t_{4}}}^{11}}}}{{3^{4}\cdot 5^{2}\cdot 11}}}+{
\frac{{{{{t_{3}}}^{3}}\,{{{t_{4}}}^{13}}}}{{2^{2}\cdot 3^{6}\cdot 5^{2}}}}+{
\frac{{2\,{{{t_{3}}}^{2}}\,{{{t_{4}}}^{19}}}}{{3^{8}\cdot 5^{3}\cdot 19}}}+{
\frac{{32\,{{{t_{4}}}^{31}}}}{{3^{13}\cdot 5^{6}\cdot 29\cdot 31}}};
\end{equation*}
\begin{equation*}
f=\frac{a}{16}C ^{4} +{\frac { {a}^{7}}{135}}C ^{3}
+
\frac {  {a}^{3} b } {6}C   ^{2}
+{\frac {
 {a}^{13}  }{2^2\cdot 3^5\cdot 5^2}}C   ^{2}
 -\frac{{a}^{5}{b}^{2}}{15},
\end{equation*}
where  $C$ is defined by the cubic   equation
$
\frac{a}{12}C ^{3}+{\frac { {a}^{7}}{90}} C ^{2}+{\frac {{a}^{3} b    }{3}}C
 +{\frac {  {a}^{13}  }{2\cdot 3^5\cdot 5^2}}C + \frac {{a}^{9}  b  }{3^4 \cdot 5} +\frac {  4\,{a}^{19} }{3^8 \cdot 5^3 \cdot 19}=c.
$ 
Swapping $t_2$ and $t_3$ gives a rational density $f$:
\medskip 
\begin{equation*}
F={\frac{1}{2}}t_{1}^{2}\,t_{4}+t_{1}\,t_{2}\,t_{3}+{\frac{{{2\,{{t_{3}^{3}}}
}\,{t_{4}}}}{{3}}}+{\frac{{{{{t_{2}^{5}}}}\,{t_{4}}}}{{240}}}+{\frac{{{
t_{2}^{3}}\,{{{t_{3}}}}\,{{{t_{4}^{3}}}}}}{{18}}}+{\frac{{{{{t_{2}}}}\,{
t_{3}^{2}}\,{{{t_{4}^{5}}}}}}{{15}}}+{\frac{{{{{t_{2}^{4}}}}\,{{{t_{4}^{7}}}}
}}{{2^{3}\cdot 3^{3}\cdot 5}}}
\end{equation*}
\begin{equation*}
+{\frac{{{t_{2}^{2}}\,{{{t_{3}}}}\,{{{t_{4}^{9}}}}}}{{2\cdot 3^{4}\cdot 5}}}+
{\frac{{8\,{{{t_{3}^{2}}}}\,{{{t_{4}^{11}}}}}}{{3^{4}\cdot 5^{2}\cdot 11}}}+{
\frac{{{{{t_{2}^{3}}}}\,{{{t_{4}^{13}}}}}}{{2^{2}\cdot 3^{6}\cdot 5^{2}}}}+{
\frac{{2\,{{{t_{2}^{2}}}}\,{{{t_{4}^{19}}}}}}{{3^{8}\cdot 5^{3}\cdot 19}}}+{
\frac{{32\,{{{t_{4}^{31}}}}}}{{3^{13}\cdot 5^{6}\cdot 29\cdot 31}}};
\end{equation*}
\begin{equation*}
f={\frac {32\,{a}^{21}}{3^8\cdot 5^4 \cdot 11^2}}+{\frac {8\,b{a}^{15}}{3^5\cdot 5^3\cdot 11}}-{
\frac {4\,c{a}^{10}}{5^2\cdot 3^4 \cdot 11}}+{\frac {2\,{b}^{2}{a}^{9}}{3^4\cdot 5^2}}-\frac{1}{30}\,b
c{a}^{4}-\frac{1}{18}\,{b}^{3}{a}^{3}+{\frac {{c}^{2}}{8\, a}}.
\end{equation*}
\medskip

\noindent {\bf Non-polynomial prepotentials}   (\ref{prep}) associated with extended affine Weyl groups can be found in \cite{DSZZ}:
 \begin{eqnarray*}
&&F=\frac{1}{2}\,{t_{{1}}^{2}}t_{{4}}+\,t_{{1}}t_{{2}}t_{{3}}-{\frac{1}{12}}
\,{t_{{2}}^{2}t_{{3}}^{2}}+{\frac{1}{720}}\,t_{{2}}{t_{{3}}^{5}}-{\frac{1}{
36288}}\,{t_{{3}}^{8}} +2t_{{2}}t_{{3}}{e^{t_{{4}}}}+\frac{1}{6}\,{t_{{3}}^{4}e^{t_{{4}}}}+
\frac{1}{2}\,e^{2t_{4}}+{\frac{1}{6}}\,{\dfrac{{t_{{2}}^{3}}}{t_{{3}}}};
\end{eqnarray*}
\begin{equation*}
f=-{\frac { C  ^{7}}{756}}+\frac{1}{48}\,b
 C  ^{4}+\frac{4}{3}\,{{\rm e}^{a}}
 C  ^{3}-2\,b{{\rm e}^{a}}+{
\frac {{b}^{3}}{2\, C  ^{2}}},
\end{equation*}
where 
$
-{\frac { C  ^{6}}{648}}+\frac{1}{36}\,b
 C  ^{3}+2\, C ^{2}{{\rm e}^{a}}-\frac{1}{6}\,{b}^{2}+{\frac {{b}^{3
}}{3 C  ^{3}}}=c.
$
Swapping $t_2$ and $t_3$ gives:
\begin{eqnarray*}
&&F=\frac{1}{2}\,{t_{{1}}^{2}}t_{{4}}+\,t_{{1}}t_{{2}}t_{{3}}-{\frac{1}{12}}
\,{t_{{2}}^{2}t_{{3}}^{2}}+{\frac{1}{720}}\,t_{{2}}^{5}{t_{{3}}}-{\frac{1}{
36288}}\,{t_{{2}}^{8}} +2t_{{2}}t_{{3}}{e^{t_{{4}}}}+\frac{1}{6}\,{t_{{2}}^{4}e^{t_{{4}}}}+
\frac{1}{2}\,e^{2t_{4}}+{\frac{1}{6}}\,{\dfrac{{t_{{3}}^{3}}}{t_{{2}}}};
\end{eqnarray*}
\begin{equation*}
f={\frac {{b}^{5}}{80}}+\frac{1}{6}\,c{b}^{3}+\frac{1}{2}\,{c}^{2} b-2\,{\rm e}^{a}b.
\end{equation*}

\medskip

\noindent {\bf Modular  prepotentials} \cite{Bertola, MS}  give rise to modular Lagrangian densities (as an example we took prepotential 4.2.2. from \cite{MS}):
$$
F=\frac{1}{2}\,{t_{{1}}^{2}}t_{{4}}+\,t_{{1}}t_{{2}}t_{{3}}-\frac{1}{4}t_2^2t_3^2\gamma(t_4)+t_2^6g_4(t_4)+t_2^4t_3g_3(t_4)+t_3^3g_1(t_4);
$$
$$
f=\frac{1}{12g_1(a)}[c+\frac{1}{2}b^2\gamma (a)]^2-g_3(a)b^4.
$$
Swapping $t_2$ and $t_3$ gives:
$$
F=\frac{1}{2}\,{t_{{1}}^{2}}t_{{4}}+\,t_{{1}}t_{{2}}t_{{3}}-\frac{1}{4}t_2^2t_3^2\gamma(t_4)+t_3^6g_4(t_4)+t_3^4t_2g_3(t_4)+t_2^3g_1(t_4);
$$
$$
f=24 C^5 g_4 (a)+8 bC^3g_3(a),
$$
where $C$ is defined by the algebraic equation
$$
30C^4 g_4+12bC^2 g_3-\frac{1}{2}b^2\gamma =c.
$$
Here $g_3=Kg_1^3, \ g_4=\frac{Kg_1}{30}(g_1'-\frac{1}{2}g_1\gamma)$
where the functions of $\gamma(a)$ and $ g_1(a)$ satisfy the ODEs
$$
\gamma'=\frac{1}{2}\gamma^2-72Kg_1^4, \quad g_1''=2\gamma g_1'-g_1\gamma',
$$
 $K=const$. The above ODE system falls within Bureau's class and its solutions are given in terms of the Schwarzian triangle functions \cite{MS}.

\subsubsection{WDVV prepotentials from integrable Lagrangian densities}
\label{sec:LWDVV}

 In view of the correspondence between integrable Lagrangians  and WDVV prepotentials 
 \begin{equation*}
F(t_{1},t_{2},t_{3},t_{4})=\frac{1}{2}t_{1}^{2}t_{4}+t_{1}t_{2}t_{3}+W(t_{2},t_{3},t_{4})
\end{equation*}
 described in Section \ref{sec:i0},  integrable Lagrangian densities  $f(a, b, c)$ constructed in this paper  give rise to prepotentials some of which are apparently new. Here we list some examples (omitting details of calculations; we will only present the corresponding function $W$).
 
 \medskip
 
\noindent{\bf Example 1.} The polynomial Lagrangian density  from Section \ref{sec:int2D},
\begin{equation*}
f=b(a^2-c^2), 
\end{equation*}
  gives rise to the  prepotential
$$
W=\frac{1}{15}t_4^{5}-t_4^{2}t_2t_3+\frac{1}{3}t_4t_2^{4}-\frac{t_3^{3}}{12t_2}.
$$

\medskip
 
\noindent{\bf Example 2.} Lagrangian densities from Section \ref{sec:di} (linear case): 
the density 
\begin{equation*}
f=\alpha e^{-b}+(e^{a}+e^{c})e^b
\end{equation*}
gives rise to the  prepotential
$$
W=-\alpha e^{t_4}-\alpha e^{-t_2}t_3-e^{t_4}e^{t_2}t_3-\frac{1}{2}t_2t_3^{2}+\frac{t_3^{2}}{2}\ln t_3; 
$$
the  density
$$
f=\alpha q(b)+(e^a+e^c)\sinh b
$$
 gives rise to the  prepotential
\begin{equation*}
W=\frac{1}{8}e^{2t_4}-e^{t_4}t_3\sinh t_2-\alpha e^{t_4}-\alpha q(t_2)t_3+\frac{1}{2}t_3^{2}\ln 
\frac{t_3}{\sinh t_2}-\frac{3}{4}t_3^{2}.
\end{equation*}
Here 
$$
q(t_2)=Li_2(-e^{t_2})-Li_2(e^{t_2})
$$
where $Li_2$ is the dilogarithm function.

\medskip

 \noindent{\bf Example 3.} Lagrangian densities from Section \ref{sec:di} (quadratic case): the density 
$$
f=e^{kb+a+c}
$$
gives rise to the prepotential
$$
W=-\frac{1}{2}t_4t_3^{2}-\frac{k}{2}t_2t_3^{2}+\frac{t_3^{2}}{2}\ln t_3;
$$
the density
$$
f=  e^{\frac{4}{\sqrt  3}b+a+c}+ e^{\frac{2}{\sqrt 3}b+2c}
$$
gives rise to the prepotential
$$
W=\frac{t_3^{2}}{2}\ln t_3-\frac{1}{2}t_4t_3^{2}-\frac{2}{\sqrt{3}}t_2t_3^{2}-\gamma e^{\frac{2}{\sqrt{3}}t_2+2t_4}t_3;
$$
the density
$$
f=\alpha e^{-\frac{1}{\sqrt2}b}+\alpha e^{ \frac{1}{\sqrt2}b+a }+e^{\frac{{1}}{\sqrt 2}b+c}+e^{\frac{3}{\sqrt 2}b+a+c}
$$
gives rise to the prepotential
$$
W=\frac{t_3^2}{2}\ln \frac{t_3}{ e^{\frac{1}{\sqrt 2}t_2}+ e^{t_4+\frac{3}{\sqrt 2}t_2}}-\frac{\alpha^2}{2}e^{t_4}-\alpha t_3\frac{1  +2 e^{t_4+{\sqrt 2}t_2}+  e^{2t_4+2\sqrt 2t_2}}{ e^{\frac{1}{\sqrt 2}t_2}+e^{t_4+\frac{3}{\sqrt 2}t_2}}.
$$

\medskip

 \noindent{\bf Example 4.} The Lagrangian density $f=e^{c}g(b, a)$ from Section \ref{sec:Jac} gives rise to the prepotential (recall that system (10) is invariant
under the interchange $a\leftrightarrow c$; for our convenience we
choose  $f=e^{c}g(b,a)$ instead of $f=e^{a}g(b,c)$):
$$
W=\frac{t_3^{2}}{2}\ln \frac{t_3}{g(t_2, t_4)}.
$$
Here
$$
g(t_2, t_4)=[\Delta(it_4/\pi)]^{-1/8}\theta_1(t_2, it_4/\pi)
$$
where $\Delta$ is the modular discriminant and $\theta_1$ is the Jacobi theta function. Note the formula
$\Delta^{1/8}(it_4/\pi)=\sqrt{2\pi^3}\ \theta_1'(0, it_4/\pi)$ where prime denotes derivative by $t_2$. The corresponding solution of WDVV equations is related to Whitham averaged one-phase solutions of NLS/Toda equations  \cite{Dubrovin}, see also \cite{Almeida, Sh}.

\medskip

 \noindent{\bf Example 5.}  The Lagrangian density
$f=\sqrt{b(\alpha a+\beta b)(\alpha b+\beta c)}$ from Section \ref{sec:trav}
gives rise to the prepotential
\begin{equation*}
W=-\frac{\beta }{4}t_{2}(\alpha t_{4}+\beta t_{2})\ln t_{3}-\frac{\alpha }{%
2\beta }t_{2}t_{3}^{2}+\frac{\beta ^{2}t_{2}^{2}}{8}\ln t_{2}+\frac{1}{8}%
(\alpha t_{4}+\beta t_{2})^{2}\ln (\alpha t_{4}+\beta t_{2}).
\end{equation*}

\section{Integrable Lagrangians in 3D}
\label{sec:int3D}

In this section we consider second-order integrable Lagrangians of the form (\ref{3}),
$$
\int f(u_{11}, u_{12}, u_{22}, u_{13}, u_{23}, u_{33}) \ dx_1dx_2dx_3,
$$
here $u_{ij}=u_{x_ix_j}$.

\subsection{Integrability conditions}
\label{sec:int3}

Let us require that all travelling wave reductions of a 3D Lagrangian density to two dimensions are integrable in the sense of Sections \ref{sec:i2} and \ref{sec:RC}. This gives the necessary conditions for integrability which, in our particular case, prove to be also sufficient. The computational details are as follows. Consider a traveling wave reduction of a 3D Lagrangian density $f(u_{11}, u_{12}, u_{22}, u_{13}, u_{23}, u_{33})$  obtained by setting $u(x_1, x_2, x_3)=v(x, y)+Q$ where $x =s_1x_1+s_3 x_3, \ y=s_2x_2+s_3 x_3$, $s_i=const$, and $Q$ is an arbitrary homogeneous quadratic form in $x_1, x_2, x_3$. We have
$$
\begin{array}{c}
u_{11}=s_1^2v_{xx}+\zeta_1, ~~u_{12}=s_1s_2v_{xy}+\zeta_2, ~~ u_{22}=s_2^2v_{yy}+\zeta_3,\\
 u_{13}=s_1s_3(v_{xx}+v_{xy})+\zeta_4, ~~
u_{23}=s_2s_3(v_{xy}+v_{yy})+\zeta_5, ~~ u_{33}=s_3^2(v_{xx}+2 v_{xy}+ v_{yy})+\zeta_6,
\end{array}
$$
where $\zeta_i$ are the coefficients of the quadratic form $Q$. Setting $v_{xx}={ a}, \ v_{xy}={ b}, \ v_{yy}={ c}$
we obtain the reduced 2D Lagrangian density ${ f}$ in the form
$$
\begin{array}{c}
{f}(a, b, c)= f(u_{11}, u_{12}, u_{22}, u_{13}, u_{23}, u_{33})\\
\ \\
=f(s_1^2 a+\zeta_1, \ s_1s_2 b+\zeta_2,\ s_2^2 c+\zeta_3,\ s_1s_3(a+b)+\zeta_4, \ s_2s_3(b+c)+\zeta_5,\ s_3^2(a+2b+c)+\zeta_6).
\end{array}
$$
We have the following differentiation rules:
\begin{equation}\label{der}
\begin{array}{c}
\partial_a=s_1^2\partial_{u_{11}}+s_1s_3\partial_{u_{13}}+s_3^2\partial_{u_{33}},\\
\ \\
\partial_b=s_1s_2\partial_{u_{12}}+s_1s_3\partial_{u_{13}}+s_2s_3\partial_{u_{23}}+ 2s_3^2\partial_{u_{33}},\\
\ \\
\partial_c=s_2^2\partial_{u_{22}}+s_2s_3\partial_{u_{23}}+s_3^2\partial_{u_{33}}.
\end{array}
\end{equation}
etc. Substituting partial derivatives of the reduced density $ f(a, b, c)$  into the 2D integrability conditions (\ref{int}) we obtain homogeneous polynomials of degree ten in $s_1, s_2, s_3$ whose coefficients are expressed in terms of partial derivatives of the original 3D density $f(u_{ij})$. Equating to zero the coefficients of these polynomials we obtain  3D integrability conditions for $f$ (note that due to the presence of arbitrary constants $\zeta_i$ the arguments of $f$  can be viewed as independent of  $s_1, s_2, s_3$). The integrability conditions can be represented in compact Hirota-type form analogous to (\ref{RC}): 
\begin{equation}\label{RC3}
(P_1Y_1v-2P_0Y_2v)[f(u_{ij})\cdot f(\tilde u_{ij})]\biggr \rvert_
{\tilde u_{ij}=u_{ij}}=0.
\end{equation}
Here the operators on the left-hand side of (\ref{RC3}) are identical to that from Section \ref{sec:RC}, with the only difference that we substitute expressions (\ref{der}) (and their tilded versions) 
for $\partial_a, \partial_b, \partial_c$ and $\partial_{\tilde a}, \partial_{\tilde b}, \partial_{\tilde c}$. Thus,
$$
\partial_a=s_1^2\partial_{u_{11}}+s_1s_3\partial_{u_{13}}+s_3^2\partial_{u_{33}}, \quad
\partial_{\tilde a}=s_1^2\partial_{\tilde u_{11}}+s_1s_3\partial_{\tilde u_{13}}+s_3^2\partial_{\tilde u_{33}},
$$
etc. The left-hand side of (\ref{RC3}) is an ${\rm Sp}(6)$-invariant operation which 
transforms a function $f$ defined on the  space of $3\times 3$ symmetric matrices $u_{ij}$ into a homogeneous form of degree four in $\xi_1, \xi_2$ and degree ten in $s_1, s_2, s_3$.

\subsection{Equivalence group in 3D}
\label{sec:equiv3}

Let ${\bf U}$ be the $3\times 3$ Hessian matrix of the function $u(x_1, x_2, x_3)$. Integrable Lagrangians of type (\ref{3}) are invariant under ${\rm Sp}(6)$-symmetry
\begin{equation}  \label{sp6}
{\bf U}\to ({\bf AU+B})({\bf CU+D})^{-1}, \quad \quad f\to \frac{f}{\det (\mathbf{CU}+\mathbf{D})},
\end{equation}
where the matrix
\begin{equation*}
\left(
\begin{array}{cc}
{\bf A} & {\bf B} \\
{\bf C} & {\bf D}
\end{array}
\right)
\end{equation*}
belongs to the symplectic group ${\rm Sp}(6, \mathbb{R})$ (here ${\bf A, B, C, D}$ are $3\times 3$
matrices). 
Note that symmetry (\ref{sp6}) suggests a relation to genus three Siegel modular forms. 
Furthermore, integrable Lagrangians (\ref{2}) are invariant under rescalings
of $f$, as well as under the addition of a `null-Lagrangian', namely,
transformations of the form
\begin{equation}  \label{null3}
f\to \lambda_0 f+\sum \lambda_{\sigma}U_{\sigma}
\end{equation}
where $U_{\sigma}$ denote all possible minors of the Hessian matrix ${\bf U}$. Transformations (\ref{sp6}) and (\ref{null3}) generate a group of dimension $21+15=36$ which
preserves the class of integrable Lagrangians (\ref{3}). 

\subsection{Examples of integrable Lagrangians in 3D}
\label{sec:ex3}

In this section we give some explicit examples of 3D integrable Lagrangian
densities $f$.

\subsubsection{Integrable Lagrangians associated with the dKP hierarchy}
\label{sec:dKP}

Here we construct three explicit  integrable Lagrangian densities arising in the context of the dKP hierarchy:
\begin{equation}\label{dKP1}
f=u_{yy}^{2}-u_{xx}u_{xt}+u_{xx}^{2}u_{yy}+u_{xx}u_{xy}^{2}+\frac{1}{4}u_{xx}^{4},
\end{equation}
\begin{equation}\label{dKP2}
f=  (u_{xy}-u_{tt}-u_{xx}u_{xt}+\frac{1}{3}u_{xx}^{3}) ^{3/2},
\end{equation}
\begin{equation}\label{dKP3}
f=u_{xt}^{-2}(u_{xt}u _{yt}-u_{xx}u_{xt}^2)^{3/2}.
\end{equation}
These examples come from the following dKP flows.

\medskip
\noindent{\bf Case 1.} The fifth-order flow of the dKP hierarchy comes from the dispersionless Lax representation
\begin{equation*}
p_{y}=\left( \frac{p^{2}}{2}+w\right) _{x}, \quad p_{t }=\left( \frac{
p^{5}}{5}+wp^{3}+vp^{2}+bp+c\right) _{x},
\end{equation*}
which gives rise to the equations
\begin{equation*}
w_{y}=v_{x},\text{ \ }b_{x}=v_{y}+3ww_{x},\text{ \ }c_{x}=b_{y}+2vw_{x},
\text{ \ }w_{t }=bw_{x}+c_{y}.
\end{equation*}
Setting $w=u_{xx},\, v=u_{xy}$ and $ b=u_{yy}+\frac{3}{2}u_{xx}^{2}$ we obtain two equations for $c$,
\begin{equation*}
c_{x}=u_{yyy}+3u_{xx}u_{xxy}+2u_{xy}u_{xxx}, \quad c_{y}=u_{xxt }-u_{yy}u_{xxx}-\frac{3}{2}
u_{xx}^{2}u_{xxx},
\end{equation*}
whose compatibility condition results in the following fourth-order PDE for $u$:
\begin{equation*}
u_{yyyy}-u_{xxxt }+3u_{xx}u_{xxyy}+2u_{xy}u_{xxxy}+3u_{xxy}^{2}+3u_{xxx}u_{xyy}+u_{yy}u_{xxxx}+\frac{3}{2}u_{xx}^{2}u_{xxxx}+3u_{xx}u_{xxx}^{2}=0.
\end{equation*}
This  is the Euler-Lagrange equation corresponding to the polynomial Lagrangian density (\ref{dKP1}).
Note that the two-dimensional  density (\ref{dKP}) is just the stationary ($t$-independent) reduction of (\ref{dKP1}).

\medskip

\noindent{\bf Case 2.} Another flow of the dKP hierarchy is associated with the Lax representation
\begin{equation*}
p_{t}=\left( \frac{p^{3}}{3}+wp+v\right) _{x},\text{ \ }p_{y}=\left( \frac{
p^{5}}{5}+wp^{3}+vp^{2}+bp+c\right) _{x},
\end{equation*}
which gives rise to the equations
\begin{equation*}
b_{t}=-bw_{x}+wb_{x}-2vv_{x}+w_{y},\quad c_{t}=wc_{x}-bv_{x}+v_{y},
\end{equation*}
\begin{equation*}
w_{t}=-2ww_{x}+b_{x},\quad v_{t}=-2vw_{x}-2wv_{x}+c_{x}.
\end{equation*}
Setting $w=u_{xx}$, $b=u_{xt}+u_{xx}^{2}$ and
$v^{2}=u_{xy}-u_{tt}-u_{xx}u_{xt}+\frac{1}{3}u_{xx}^{3}$
we obtain two equations for $c$,
\begin{equation*}
c_{t}=v_{y}+u_{xx}v_{t}+2vu_{xx}u_{xxx}+(u_{xx}^{2}-u_{xt})v_{x},\text{ \ }c_{x}=v_{t}+2vu_{xxx}+2u_{xx}v_{x},
\end{equation*}
whose compatibility condition yields
\begin{equation*}
v_{tt}+(vu_{xx})_{xt}=v_{xy}+(v(u_{xx}^{2}-u_{xt}))_{xx}.
\end{equation*}
This PDE is the Euler-Lagrange equation corresponding to  the  density (\ref{dKP2}).

\medskip

\noindent{\bf Case 3.} This example comes from the dispersionless Lax pair
\begin{equation*}
p_{t}=\left( \frac{r}{p-q}\right) _{x},\text{ \ }p_{y}=\left( \frac{p^{3}}{3}
+wp+v\right) _{x},
\end{equation*}
which gives rise to the equations
\begin{equation*}
q_{y}=q^{2}q_{x}+qw_{x}+wq_{x}+v_{x},
\quad
r_{y}=q^{2}r_{x}+2qrq_{x}+rw_{x}+wr_{x},
\quad
w_{t}=-r_{x},
\quad
v_{t}=-qr_{x}-rq_{x}.
\end{equation*}
Setting $w=u_{xx}$,  $r=-u _{xt}$, $ q^{2}=\frac{u_{yt}}{u_{xt}}-u_{xx}$, the second and the third equations will be satisfied identically, while the first and the fourth imply  
\begin{equation*}
v_{x}=q_{y}-\left( \frac{1}{3}q^{3}+qu_{xx}\right) _{x}, \quad v_{t}=(u_{xt}q)_{x},
\end{equation*}
whose consistency condition gives
\begin{equation*}
q_{yt}-\left[ \frac{1}{3}q\left( \frac{u_{yt}}{u_{xt}}+2u_{xx}\right) \right] _{xt}=(u_{xt}q)_{xx}.
\end{equation*}
This is the Euler-Lagrange equation corresponding to the  density (\ref{dKP3}).

\subsubsection{Integrable Lagrangian densities of the form $f=f(u_{xy},u_{xt},u_{yt})$}
\label{sec:Lob}

We will show that the general integrable density of this form  is expressible in terms of the Lobachevsky function $L(s)=-\int_0^s\ln  \cos \xi \ d\xi$, a special function which features in Lobachevsky's formulae for hyperbolic volumes \cite{Lob}. 

Using the notation $u_{xy}=v_3, \ u_{xt}=v_2, \ u_{yt}=v_1$ and $f_{ij}=f_{v_iv_j}$ one can show that the integrability conditions (which we do not present here explicitly) can be rewritten as simple relations for the $2\times 2$ minors of the Hessian matrix  $F=Hess f$. Namely, they are equivalent to the conditions that the minors  
$$
f_{11}f_{22}-f_{12}^2=a_3, \quad  f_{11}f_{33}-f_{13}^2=a_2, \quad  f_{22}f_{33}-f_{23}^2=a_1, 
$$
and
$$
f_{12}f_{13}-f_{11}f_{23}=p_1, \quad  f_{12}f_{23}-f_{22}f_{13}=p_2, \quad  f_{13}f_{23}-f_{33}f_{12}=p_3, 
$$
are such that $a_i=const$ and $p_i$ is a function of the argument $v_i$ only. This gives the inverse of the Hessian matrix $F$  in the form
\begin{equation}\label{F}
\left(
\begin{array}{ccc}
f_{11}&f_{12}&f_{13}\\
f_{12}&f_{22}&f_{23}\\
f_{13}&f_{23}&f_{33}\\
\end{array}
\right)^{-1}=\frac{1}{\det F}
\left(
\begin{array}{ccc}
a_1&p_3&p_2\\
p_3&a_2&p_1\\
p_2&p_1&a_3
\end{array}
\right).
\end{equation}
Taking the determinant of both sides we obtain
\begin{equation}\label{det}
\det F=\sqrt{a_1a_2a_3-a_1p_1^2-a_2p_2^2-a_3p_3^2+2p_1p_2p_3}.
\end{equation}
Inverting the matrix identity (\ref{F}) gives
\begin{equation}\label{F1}
\left(
\begin{array}{ccc}
f_{11}&f_{12}&f_{13}\\
f_{12}&f_{22}&f_{23}\\
f_{13}&f_{23}&f_{33}\\
\end{array}
\right)=\frac{1}{\det F}
\left(
\begin{array}{ccc}
a_2a_3-p_1^2&p_1p_2-a_3p_3&p_1p_3-a_2p_2\\
p_1p_2-a_3p_3&a_1a_3-p_2^2&p_2p_3-a_1p_1\\
p_1p_3-a_2p_2&p_2p_3-a_1p_1&a_1a_2-p_3^2
\end{array}
\right)
\end{equation}
where we use (\ref{det}) for $\det F$. The consistency conditions of equations (\ref{F1}) lead to simple  ODEs for the functions $p_i(v_i)$:
$$
p_1'=c(p_1^2-a_2a_3), \quad p_2'=c(p_2^2-a_1a_3), \quad p_3'=c(p_3^2-a_1a_2),
$$
where $c$ is yet another arbitrary constant. The further analysis depends on how many constants among $a_i$ are equal to zero.

\noindent{\bf All constants  are zero.} In this case without any loss of generality one can set $p_i=1/v_i$ which leads to the integrable Lagrangian density
$$
f=\sqrt{u_{xy}u_{xt}u_{yt}}.
$$

\noindent{\bf Two constants  are zero.} Then  one can also set $p_i=1/v_i$. Modulo (complex) rescalings this leads to the Lagrangian density
\begin{equation*}
\begin{array}{c}
f=\sqrt{u_{xy}u_{xt}(2u_{yt}-u_{xy}u_{xt})}-2u_{yt}\arctan \sqrt{\frac{2u_{yt}}{u_{xy}u_{xt}}-1}.
\end{array}
\end{equation*}

\noindent{\bf One constant is zero.} This leads to the Lagrangian density
\begin{equation*}
\begin{array}{c}
f=(u_{xt}-u_{xy})\arctan \frac{\sqrt{2u_{xt}u_{xy}\coth u_{yt}-u_{xt}^{2}-u_{xy}^{2}%
}}{u_{xt}-u_{xy}}-(u_{xt}+u_{xy})\arctan \frac{\sqrt{2u_{xt}u_{xy}\coth
u_{yt}-u_{xt}^{2}-u_{xy}^{2}}}{u_{xt}+u_{xy}}.
\end{array}
\end{equation*}

\noindent{\bf All constants  are nonzero.} This case  is more interesting. Setting $a_i=1, \ c=-1$ we obtain 
$p_i'=1-p_i^2$ so that $p_i=\tanh v_i$. Equations (\ref{F1}) can be integrated once to yield
$$
f_1=\arcsin \frac{p_1-p_2p_3}{\sqrt {1-p_2^2}\sqrt {1-p_3^2}}, \quad
f_2=\arcsin \frac{p_2-p_1p_3}{\sqrt {1-p_1^2}\sqrt {1-p_3^2}}, \quad
f_3=\arcsin \frac{p_3-p_1p_2}{\sqrt {1-p_1^2}\sqrt {1-p_2^2}}. 
$$
Choosing $p_i$ as the new independent variables  we obtain
\begin{equation*}
\begin{array}{c}
f_{p_1}=\frac{1}{1-p_1^2}\arcsin \frac{p_1-p_2p_3}{\sqrt {1-p_2^2}\sqrt {1-p_3^2}}, \\
\ \\
f_{p_2}=\frac{1}{1-p_2^2}\arcsin \frac{p_2-p_1p_3}{\sqrt {1-p_1^2}\sqrt {1-p_3^2}}, \\
\ \\
f_{p_3}=\frac{1}{1-p_3^2}\arcsin \frac{p_3-p_1p_2}{\sqrt {1-p_1^2}\sqrt {1-p_2^2}}, 
\end{array}
\end{equation*}
or, in differentials, 
\begin{equation}\label{f_p}
\begin{array}{c}
df=\frac{dp_1}{1-p_{1}^{2}}\arcsin \frac{p_{1}-p_{2}p_{3}}{\sqrt{1-p_{2}^{2}}
\sqrt{1-p_{3}^{2}}}+
\frac{dp_2}{1-p_{2}^{2}}\arcsin \frac{p_{2}-p_{1}p_{3}
}{\sqrt{1-p_{1}^{2}}\sqrt{1-p_{3}^{2}}}\\
\ \\
+\frac{dp_3}{1-p_{3}^{2}}\arcsin 
\frac{p_{3}-p_{1}p_{2}}{\sqrt{1-p_{1}^{2}}\sqrt{1-p_{2}^{2}}}.
\end{array}
\end{equation}
In the original variables $v_1, v_2, v_3$ relation (\ref{f_p}) takes the form 
\begin{equation*}
\begin{array}{c}
df=\arcsin(\cosh v_{2}\cosh v_{3}\tanh
v_{1}-\sinh v_{2}\sinh v_{3})\ dv_1\\
\ \\+\arcsin(\cosh v_{1}\cosh v_{3}\tanh
v_{2}-\sinh v_{1}\sinh v_{3})\ dv_2\\
\ \\
+\arcsin(\cosh v_{1}\cosh v_{2}\tanh
v_{3}-\sinh v_{1}\sinh v_{2})\ dv_3.
\end{array}
\end{equation*}


\medskip

\noindent{\bf Remark 9.} Relation (\ref{f_p}) has an unexpected link to spherical trigonometry. 
On the unit sphere $S^2$, consider a spherical triangle $\triangle ABC$ with interior angles $A, B, C$ and side lengths $a, b, c$ (so that side $a$ lies opposite the angle $A$, etc). The spherical laws of cosines are 
\begin{equation}\label{sl}
\begin{array}{c}
\cos a=\cos b\cos c+\sin b\sin c\cos A,\\
\cos b=\cos a\cos c+\sin a\sin c\cos B,\\
\cos c=\cos a\cos b+\sin a\sin b\cos C,
\end{array}
\end{equation}
and 
\begin{equation}\label{sl1}
\begin{array}{c}
\cos A=-\cos B\cos C+\sin B\sin C\cos a,\\
\cos B=-\cos A\cos C+\sin A\sin C\cos b,\\
\cos C=-\cos A\cos B+\sin A\sin B\cos c,
\end{array}
\end{equation}
respectively.  Note that the map $(A, B, C) \to (a, b, c)$ sending  angles of a spherical triangle  to its side lengths  is integrable in the sense of multidimensional consistency \cite{PS}, and is closely related to the discrete Darboux system \cite{BK, KS}. 
Setting 
\begin{equation} \label{pi}
p_{1}=\cos a,\text{ \ }p_{2}=\cos b,\text{ \ }p_{3}=\cos c
\end{equation}
and using (\ref{sl}) we can rewrite  (\ref{f_p})  in the following  Schl\"afly-type form:
\begin{equation}\label{S}
df=(A-\pi/2)\frac{da}{\sin a}+(B-\pi/2)\frac{db}{\sin b}+(C-\pi/2)\frac{dc}{\sin c}.
\end{equation}
Recall that the classical Schl\"afly formula expresses the differential of the volume of a spherical polyhedron in terms of its side lengths and dihedral angles. Expression (\ref{S}), which can be viewed as a two-dimensional Schl\"afly formula, has appeared in \cite{Luo, Luo1} as a special  case of a one-parameter family of closed Schl\"afly-type forms associated with spherical triangles (case $h=0$ of Theorem 3.2(b) in \cite{Luo1}).  Note that the function $f$ of a spherical triangle defined by (\ref{S}) is essentially the volume of the ideal hyperbolic octahedron which is the convex hull of the six intersection points of the three circles on the sphere at infinity bounding the spherical triangle  $\triangle ABC$
(\cite{Luo1}, Appendix C, see also \cite{Luo2}). This function $f$ is related to the `capacity' function of a spherical triangle.  Expressions similar to (\ref{S}) have appeared before in the context of variational principles for circle packings and triangulated surfaces \cite{CdV, Bobenko, Bobenko1, Luo}.
\medskip

Integration of  (\ref{f_p}) is quite non-trivial, the answer is given in terms of the Lobachevsky function. The computations below are based on  formula (\ref{S}) and the spherical cosine laws (\ref{sl}), (\ref{sl1}).  Using
$\frac{da}{\sin a}=\frac{1}{2}d\ln  \frac{1-\cos a}{1+\cos a}$ we can rewrite $df$ in the form
$$
\begin{array}{c}
df=-\frac{\pi }{4}d\left( \ln  \frac{1-\cos a}{1+\cos a}+\ln \frac{1-\cos b}{
1+\cos b}+\ln \frac{1-\cos c}{1+\cos c}\right)\\
\ \\
 +\frac{A}{2}d\ln \frac{1-\cos a}{1+\cos a}+\frac{B}{2}d\ln \frac{1-\cos b}{1+\cos b}+\frac{C}{2}d\ln 
\frac{1-\cos c}{1+\cos c}.
\end{array}
$$
Equivalently,
$$
\begin{array}{c}
df=-\frac{\pi }{4}d\left( \ln  \frac{1-\cos a}{1+\cos a}+\ln \frac{1-\cos b}{
1+\cos b}+\ln \frac{1-\cos c}{1+\cos c}\right)\\
\ \\
 +d\left(\frac{A}{2}\ln \frac{1-\cos a}{1+\cos a}+\frac{B}{2}\ln \frac{1-\cos b}{1+\cos b}+\frac{C}{2}\ln 
\frac{1-\cos c}{1+\cos c}\right)\\
\ \\
-\frac{1}{2}\left(\ln \frac{1-\cos a}{1+\cos a}dA+\ln \frac{1-\cos b}{1+\cos b}dB+\ln 
\frac{1-\cos c}{1+\cos c}dC \right).
\end{array}
$$
Let us rewrite the last term of this expression as a total differential. Note that using (\ref{sl1}) we have
\begin{equation*}
\frac{1-\cos a}{1+\cos a}=\frac{\sin B\sin C-\cos A-\cos B\cos C}{\sin B\sin C+\cos
A+\cos B\cos C}=\frac{-\cos A-\cos (B+C)}{\cos A+\cos (B-C)}=\frac{\cos \frac{2\pi-A-B-C}{2}\cos \frac{B+C-A}{2}}{\cos \frac{A+B-C}{2}\cos \frac{A+C-B}{2}}.
\end{equation*}
With similar formulae for $\frac{1-\cos b}{1+\cos b}$ and $\frac{1-\cos c}{1+\cos c}$ we obtain:
\begin{equation*}
\begin{array}{c}
-\frac{1}{2}\left(\ln \frac{1-\cos a}{1+\cos a}dA+\ln \frac{1-\cos b}{1+\cos b}dB+\ln 
\frac{1-\cos c}{1+\cos c}dC \right)\\
\ \\
=-\frac{1}{2}\ln \frac{\cos \frac{2\pi-A-B-C}{2}\cos \frac{B+C-A}{2}}{\cos \frac{A+B-C}{2}\cos \frac{A+C-B}{2}} dA-\frac{1}{2}\ln \frac{\cos \frac{2\pi-A-B-C}{2}\cos \frac{A+C-B}{2}}{\cos \frac{A+B-C}{2}\cos \frac{B+C-A}{2}}dB-
\frac{1}{2}\ln \frac{\cos \frac{2\pi-A-B-C}{2}\cos \frac{A+B-C}{2}}{\cos \frac{B+C-A}{2}\cos \frac{A+C-B}{2}}dC\\
\ \\
=\ln \cos \frac{2\pi-A-B-C}{2}d\left(\frac{2\pi-A-B-C}{2}\right)+\ln \cos \frac{A+B-C}{2}d\left(\frac{
A+B-C}{2}\right)\\
\ \\
+\ln \cos \frac{A+C-B}{2}d\left(\frac{A+C-B}{2}\right)+\ln \cos \frac{B+C-A}{2}d\left(\frac{
B+C-A}{2}\right).
\end{array}
\end{equation*}
On integration, we  obtain the final formula for $f$:
\begin{equation*}
\begin{array}{c}
f=-\frac{\pi }{4}\left( \ln  \frac{1-\cos a}{1+\cos a}+\ln \frac{1-\cos b}{
1+\cos b}+\ln \frac{1-\cos c}{1+\cos c}\right)\\
 \ \\
+\frac{A}{2}\ln \frac{1-\cos a}{1+\cos a}+\frac{B}{2}\ln \frac{1-\cos
b}{1+\cos b}+\frac{C}{2}\ln \frac{1-\cos c}{1+\cos c} \\
\ \\
-L\left(\frac{2\pi-A-B-C}{2}\right) -L\left(\frac{A+B-C}{2}\right) -L\left(\frac{A+C-B}{2}\right) -L\left(\frac{B+C-A}{2}\right), 
\end{array}
\end{equation*}
where $L(s)=-\int_0^s\ln  \cos \xi  \ d\xi$ is the Lobachevsky function. In the original variables $v_1, v_2, v_3$ defined as 
$p_1=\cos a=\tanh v_1$, $p_2=\cos b=\tanh v_2$, $p_3=\cos c=\tanh v_3$ this gives:
\begin{equation*}
\begin{array}{c}
f=\frac{\pi }{2}(v_{1}+v_{2}+v_{3})-(Av_{1}+Bv_{2}+Cv_{3})\\
\ \\
-L\left(\frac{2\pi-A-B-C}{2}\right) -L\left(\frac{A+B-C}{2}\right) -L\left(\frac{A+C-B}{2}\right) -L\left(\frac{B+C-A}{2}\right);
\end{array}
\end{equation*}
here $A, B, C$ are defined, as functions of $v_1, v_2, v_3$, via the spherical cosine laws: 
\begin{equation*}
\begin{array}{c}
\cos A=\frac{\cos a-\cos b\cos c}{\sin b\sin c}=\cosh v_{2}\cosh v_{3}\tanh
v_{1}-\sinh v_{2}\sinh v_{3},\\
\ \\
\cos B=\frac{\cos b-\cos a\cos c}{\sin a\sin c}=\cosh v_{1}\cosh v_{3}\tanh
v_{2}-\sinh v_{1}\sinh v_{3},\\
\ \\
\cos C=\frac{\cos c-\cos a\cos b}{\sin a\sin b}=\cosh v_{1}\cosh v_{2}\tanh
v_{3}-\sinh v_{1}\sinh v_{2}.
\end{array}
\end{equation*}
Note that the linear term $\frac{\pi }{2}(v_{1}+v_{2}+v_{3})$ can be ignored as it does not effect the  Euler-Lagrange equations corresponding to the density $f$.

\medskip

\noindent{\bf Euler-Lagrange equation.} In terms of the side lengths $a, b, c$ and angles $A, B, C$ of a spherical triangle,
 the Euler-Lagrange equation corresponding to the density $f$ takes the form
 \begin{equation}\label{ELg}
 \begin{array}{c}
 A_{yt}+B_{xt}+C_{xy}=0,\\
 \ \\
 \frac{a_x}{\sin a}= \frac{b_y}{\sin b}= \frac{c_t}{\sin c},
 \end{array}
 \end{equation}
where we keep in mind the spherical cosine laws (\ref{sl}), (\ref{sl1}). Indeed, the first relation is equivalent to
$(f_1)_{yt}+(f_2)_{xt}+(f_3)_{xy}=0$, while the second set of relations comes from the consistency conditions of  relations (\ref{pi}) rewritten in the form
$u_{xy}={\rm arctanh}(\cos c)$, $u_{xt}={\rm arctanh}(\cos b)$, $u_{yt}={\rm arctanh}(\cos a)$.

\medskip

\noindent{\bf Remark 10.} Similar analysis of the Lagrangian densities   $f=f(u_{xx},u_{yy},u_{tt})$ gives no interesting examples: one can show that in this case the integrability conditions 
imply that {\it all} $2\times 2$ minors of the Hessian matrix of $f$  must necessarily be constant, thus leading to quadratic 
densities $f$ with linear Euler-Lagrange equations.

\subsection{2D densities as travelling wave reductions of 3D densities}
\label{sec:trav}

Given a 3D integrable Lagrangian density $ f(u_{xx},u_{xy},u_{yy}, u_{xt}, u_{yt}, u_{tt})$ one can apply a  travelling wave ansatz, $u(x, y, t)=u(\xi, \eta)$ where $\xi=a_1x+a_2y+a_3t, \ \eta= b_1 x+ b_2 y+b_3 t$, to obtain an integrable 2D Lagrangian density of the form $f(u_{\xi \xi}, u_{\xi \eta}, u_{\eta \eta})$. In fact, modulo linear transformations of $\xi$ and $\eta$ it is sufficient to assume $\xi=x+\alpha t, \ \eta= y+\beta t$. Applying this construction to the density $f=\sqrt{u_{xy}u_{xt}u_{yt}}$ found in Section \ref{sec:Lob} one obtains 2D integrable densities of the form
$$
f=\sqrt{u_{\xi \eta}(\alpha u_{\xi \xi}+\beta u_{\xi \eta})(\alpha u_{\xi \eta}+\beta u_{\eta \eta}) }.
$$

\section{Dispersive deformations of integrable Lagrangian densities}
\label{sec:deform}

Some integrable Lagrangian densities  possess integrable dispersive deformations (both  in 2D and 3D). Here we give three examples (a complete classification of integrable dispersive deformations is a non-trivial open problem).
\medskip

\noindent{\bf Example 1.} The Lagrangian density (\ref{dKP1}), 
$$
f=u_{yy}^{2}-u_{xx}u_{xt}+u_{xx}^{2}u_{yy}+u_{xx}u_{xy}^{2}+\frac{1}{4}u_{xx}^{4},
$$
(Section \ref{sec:dKP}, case 1) possesses integrable dispersive deformation
\begin{equation*}
f_{\epsilon}=u_{yy}^{2}-u _{xx}u_{xt}+u_{xx}^{2}u_{yy}+u_{xx}u_{xy}^{2}+\frac{1}{4}u_{xx}^{4}-\epsilon ^{2}u_{xx}u_{xxx}^{2}-\frac{\epsilon ^{2}}{2}u_{xxy}^{2}+\frac{\epsilon ^{4}}{80}u_{xxxx}^{2},
\end{equation*}
here $\epsilon$ is a deformation parameter.  The corresponding (dispersive) Euler-Lagrange equation has the Lax pair
\begin{equation*}
\epsilon \psi _{y}=\frac{\epsilon ^{2}}{2}\psi _{xx}+a\psi ,\quad
\epsilon \psi _{t}=\frac{\epsilon ^{5}}{5}\psi _{xxxxx}+\epsilon ^{3}a\psi
_{xxx}+\epsilon ^{2}b\psi _{xx}+\epsilon c\psi _{x}+w\psi,
\end{equation*}
where  $a=u_{xx},\ 
b=u_{xy}+\frac{3\epsilon }{2}u_{xxx},\ c=u _{yy}+
\frac{3}{2}u_{xx}^{2}+\epsilon u _{xxy}+\frac{5\epsilon ^{2}}{4}
u_{xxxx}$, and the variable $w$ is defined by the equations
\begin{equation*}
w_{x}=u_{yyy}+3u_{xx}u _{xxy}+2u _{xy}u _{xxx}+
\frac{3\epsilon }{2}u _{xxx}^{2}+\frac{3\epsilon }{2}u _{xx}u_{xxxx}+\frac{\epsilon }{2}u _{xxyy}+\frac{3\epsilon ^{2}}{4}u_{xxxxy}+\frac{3\epsilon ^{3}}{8}u_{xxxxxx},
\end{equation*}
\begin{equation*}
w_{y}=u _{xxt}+\frac{\epsilon }{2}u _{xyyy}+\frac{\epsilon ^{2}}{4}
u_{xxxyy}+\frac{3\epsilon ^{3}}{8}u _{xxxxxy}-\frac{\epsilon ^{4}
}{80}u _{xxxxxxx}.
\end{equation*}
The stationary  reduction of Example 1 provides dispersive deformation of the    two-dimensional  density (\ref{dKP}).

\medskip

\noindent{\bf Example 2.}
The Lagrangian density (\ref{dKP2}),
\begin{equation*}
f=\left(u_{xy}-u_{tt}-u_{xx}u_{xt}+\frac{1}{3}u_{xx}^{3}\right)^{3/2},
\end{equation*}
(Section \ref{sec:dKP}, case 2) possesses integrable dispersive deformation
\begin{equation*}
f_{\epsilon}=\left(u_{xy}-u_{tt}-u_{xx}u_{xt}+\frac{1}{3}u_{xx}^{3}+\frac{\epsilon ^{2}}{12}(4u _{xx}u_{xxxx}+3u_{xxx}^{2}-4u_{xxxt})+\frac{\epsilon ^{4}}{45}u_{xxxxxx}\right)^{3/2}.
\end{equation*}
The corresponding dispersive Euler-Lagrange equation has the Lax pair
\begin{equation*}
\epsilon \psi _{t}=\frac{\epsilon ^{3}}{3}\psi _{xxx}+\epsilon w\psi
_{x}+v\psi ,\quad \epsilon \psi _{y}=\frac{\epsilon ^{5}}{5}\psi
_{xxxxx}+\epsilon ^{3}w\psi _{xxx}+\epsilon ^{2}({v+\epsilon w_{x}})\psi
_{xx}+\epsilon b\psi _{x}+c\psi, 
\end{equation*}
where
\begin{equation*}
w=u_{xx},\quad v=f_{\epsilon}^{1/3}+\frac{\epsilon }{2}u_{xxx}, \quad
b=u_{xt}+u_{xx}^{2}+\frac{2\epsilon ^{2}}{3}u_{xxxx}+\epsilon v_{x},
\end{equation*}
and the function $c$ is determined by the equations
\begin{equation*}
c_{x}=v_{t}+2(u_{xx}v)_{x}+\frac{2}{3}\epsilon ^{2}v_{xxx},
\end{equation*}
\begin{equation*}
c_{t}=v_{y}+(u_{xx}v)_{t}+\left( vu_{xx}^{2}-vu_{xt}-\epsilon vv_{x}-\frac{\epsilon ^{2}}{3}u_{xx}v_{xx}-\frac{2\epsilon ^{2}}{3}v_{x}u_{xxx}+\frac{\epsilon ^{2}}{3}c_{xx}-\frac{\epsilon ^{4}}{5}v_{xxxx}\right) _{x}.
\end{equation*}

\noindent{\bf Example 3.}
The Lagrangian density (\ref{dKP3}), 
\begin{equation*}
f=u_{xt}^{-2}\left(u_{xt}u_{yt}-u_{xx}u_{xt}^{2}\right) ^{3/2},
\end{equation*}
(Section \ref{sec:dKP}, case 3) possesses integrable dispersive deformation
\begin{equation*}
f_{\epsilon}=u_{xt}^{-2}\left(u_{xt}u_{yt}-u_{xx}u_{xt}^{2}+\frac{\epsilon ^{2}}{4}u_{xxt}^{2}-\frac{\epsilon ^{2}}{3}
u_{xt}u_{xxxt}\right) ^{3/2},
\end{equation*}
 The corresponding dispersive Euler-Lagrange equation comes from the Lax pair
\begin{equation*}
\epsilon \psi _{y}=\frac{\epsilon^3}{3}\psi _{xxx}+\epsilon w\psi _{x}+v\psi ,\quad \epsilon^2 \psi
_{xt}=\epsilon q\psi _{t}+r\psi,
\end{equation*}
where
\begin{equation*}
w=u_{xx},\ \text{\ \ }r=-u_{xt},\text{ \ }q=\left( \frac{f_{\epsilon }}{
u_{xt}}\right) ^{1/3}+\frac{\epsilon }{2}\frac{u_{xxt}}{u_{xt}},
\end{equation*}
and the variable $v$ is defined by the equations
\begin{equation*}
v_{t}=(u_{xt}q)_{x},\text{ \ }v_{x}=q_{y}-\left(u_{xx}q+\frac{1}{3}
q^{3}+\epsilon qq_{x}+\frac{\epsilon^2}{3}q_{xx}\right)_{x}.
\end{equation*}

\section{Concluding remarks}

Here we list some problems for further study.

\begin{itemize}
\item {\bf Multi-dimensional Lagrangians.} It would be of interest to describe  multi-dimensional versions of second-order integrable Lagrangians. Thus,  anti-self-dual four-manifolds with a parallel real spinor are described by the
integrable 4D Dunajski system \cite{Dun}
\begin{equation*}
a_{xt}+a_{yz}+u _{xx}a_{yy}+u_{yy}a_{xx}-2u_{xy}a_{xy}=0,
\end{equation*}
\begin{equation*}
u_{xt}+u_{yz}+u_{yy}u_{xx}-u_{xy}^{2}=a,
\end{equation*}
which can be written as a single fourth-order PDE for the function $u$. This PDE comes from  the second-order Lagrangian
\begin{equation*}
\int (u_{xt}+u_{yz}+u_{yy}u_{xx}-u_{xy}^{2})^{2}\ dxdydzdt.
\end{equation*}
Similarly, anti-self-dual scalar flat four-manifolds (Flaherty-Park spaces, see  \cite{Takasaki} and references therein) are governed by the equations
\begin{equation*}
u _{xz}(\ln F)_{yt}-u_{xy}(\ln F)_{zt}-u_{zt}(\ln
F)_{xy}+u_{yt}(\ln F)_{xz}=0,
\end{equation*}
\begin{equation*}
u_{xz}u_{yt}-u_{xy}u_{zt}=F,
\end{equation*}
which are equivalent to a single fourth-order PDE for $u$. The corresponding Lagrangian is 
\begin{equation*}
S=\int [F\ln F-F]\ dxdydzdt,
\end{equation*}
where one has to substitute $F=u_{xz}u_{yt}-u_{xy}u_{zt}$.

\item {\bf Multi-component Lagrangians.} Our approach can be generalised in a straightforward way to describe 2-field integrable Lagrangians of the form 
$$
\int f(u_x, u_y, v_x, v_y)\ dxdy,
$$
as well as their 3D analogues,
$$
\int f(u_x, u_y, u_t, v_x, v_y, v_t)\ dxdydt.
$$

\item {\bf Higher-order quasilinear PDEs.} Similarly, one can classify third-order integrable PDEs of the form
$$
a_1u_{xxx}+a_2u_{xxy}+a_3u_{xyy}+a_4u_{yyy}=0
$$
where the coefficients $a_i$ are functions of the second-order derivatives $u_{xx}, u_{xy}, u_{yy}$ only. This problem also has a natural 3D analogue.

\end{itemize}

\section*{Acknowledgments}

We thank A. Basalaev, A. Bobenko, L. Bogdanov, Yu. Brezhnev, D. Guzzetti, A. Mednykh, V. Shramchenko, I. Strachan and A. Verbovetsky for useful discussions. EVF was supported by the EPSRC grant EP/N031369/1. MVP was supported by the Russian Foundation for
Fundamental Research (grant 18-51-18007) and an LMS scheme 4 grant.  LX was supported by the
National Natural Science Foundation of China (grant numbers: 11501312 and 11775121).

\end{document}